%% file: SMP-17-005_temp.tex
\pdfoutput=1

\documentclass[11pt,twoside,a4paper,cmspaper,final,collab]{cms-tdr}
\def\svnVersion{485436P}\def\cmsCernNoTag{CERN-EP-2018-161}\def\cmsCernDate{\today}\def\cmsMessage{Published in Physics Letters B as \href{http://dx.doi.org/10.1016/j.physletb.2018.11.007}{\doi{10.1016/j.physletb.2018.11.007.}}}

\begin{document}\cmsNoteHeader{SMP-17-005}

\hyphenation{had-ron-i-za-tion}
\hyphenation{cal-or-i-me-ter}
\hyphenation{de-vices}
\RCS$HeadURL: svn+ssh://svn.cern.ch/reps/tdr2/papers/SMP-17-005/trunk/SMP-17-005.tex $
\RCS$Id: SMP-17-005.tex 482705 2018-11-28 16:57:21Z gpinnaan $

\newlength\cmsFigWidth

\ifthenelse{\boolean{cms@external}}{\setlength\cmsFigWidth{0.8\columnwidth}}{\setlength\cmsFigWidth{0.49\textwidth}}
\ifthenelse{\boolean{cms@external}}{\providecommand{\cmsLeft}{top\xspace}}{\providecommand{\cmsLeft}{left\xspace}}
\ifthenelse{\boolean{cms@external}}{\providecommand{\cmsRight}{bottom\xspace}}{\providecommand{\cmsRight}{right\xspace}}
\providecommand{\cmsTable}[1]{\resizebox{\textwidth}{!}{#1}}
\providecommand{\NA}{\ensuremath{\text{---}}}
\newlength\cmsTabSkip\setlength{\cmsTabSkip}{1ex}
\providecommand{\cmsTable}[1]{\resizebox{\textwidth}{!}{#1}}

\cmsNoteHeader{SMP-17-005}

\title{Measurement of differential cross sections for \PZ boson pair production in association with jets at $\sqrt{s}=8$ and 13\TeV}

\date{\today}

\newcommand{\MADGRAPHf} {\textsc{MadGraph5}\xspace}
\newcommand{\MGNLO}{\textsc{MG5}\_a\MCATNLO}
\newcommand{\PHANTOM}{\textsc{Phantom}\xspace}
\newcommand{\mjj}{\ensuremath{m_{\mathrm{jj}}}\xspace}
\newcommand{\detajj}{\ensuremath{\Delta\eta_{\mathrm{jj}}}\xspace}
\newcommand{\aetaj}{\ensuremath{\abs{\eta_{\mathrm{j}}}}\xspace}
\newcommand{\aetal}{\ensuremath{\abs{\eta_{\mathrm{\ell}}}}\xspace}
\newcommand{\ptl}{\ensuremath{\pt^{\ell}}\xspace}
\newcommand{\Njets}{\ensuremath{N_{\text{jets}}}\xspace}
\newcommand{\zjets}{\ensuremath{\PZ\text{+jets}}\xspace}
\newcommand{\zzjets}{\ensuremath{\PZ\PZ\text{+jets}}\xspace}
\newcommand{\wzjets}{\ensuremath{\PW\PZ\text{+jets}}\xspace}

\abstract{
This Letter reports measurements of differential cross sections for the production of two \PZ bosons in association with jets
in proton-proton collisions at $\sqrt{s} =8$ and 13\TeV.
The analysis is based on data samples collected at the LHC with the CMS detector, corresponding to integrated luminosities of 19.7 and 35.9\fbinv at 8 and 13\TeV, respectively.
The measurements are performed in the leptonic decay modes $\PZ\PZ\to\ell^+ \ell^-  \ell'^+ \ell'^-$, where $\ell,\ell' = \Pe, \mu$.
The differential cross sections as a function of the jet multiplicity, the transverse momentum \pt, and pseudorapidity of the \pt-leading and subleading jets are presented.
In addition, the differential cross sections as a function of variables sensitive to the vector boson scattering, such as the invariant mass of the two \pt-leading jets and their
pseudorapidity separation, are reported. The results are compared to theoretical predictions and found in good agreement within the theoretical and experimental uncertainties.
}

\hypersetup{
pdfauthor={CMS Collaboration},
pdftitle={Measurement of differential cross sections for Z boson pair production in association with jets at sqrt(s)=8 and 13 TeV},
pdfsubject={CMS},
pdfkeywords={CMS, physics, SM, ZZ, ZZjets, VBS}}

\maketitle

\section{Introduction}
\label{sec:introduction}
The production of massive vector boson pairs is a key process for the understanding of both the non-Abelian gauge
structure of the standard model (SM) and of the electroweak symmetry breaking mechanism.
Thus, relevant information can be gathered measuring vector boson scattering~\cite{Lemoine:1979pm} and triboson production processes
that occur through the electroweak (EW) production of jets in association with bosons.
Because of the very low cross sections for these processes compared
to others leading to the same final state, a detailed understanding of the quantum chromodynamics (QCD) corrections
to the associated production of vector boson pairs and jets is of paramount importance. The analysis presented in this Letter
has been designed to provide such detailed understanding.

Both the ATLAS and CMS Collaborations have measured the inclusive production cross section of \PZ boson pairs
and the differential cross sections as a function of \PZ boson pair observables~\cite{AtlasZZ7Tev,Aad:2015zqe,Aaboud:2017rwm,ZZXS7TeVPaper,ZZXS8TeVPaper,ZZXS13TeV2015Paper,ZZXS13TeVPaper}.
In this Letter we present new measurements of differential cross sections for the production of two \PZ bosons in association with jets in proton-proton (pp) collisions
at $\sqrt{s}=8$ and $13\TeV$ that extend the analyses of Refs.~\cite{ZZXS8TeVPaper, ZZXS13TeVPaper} to jet variables.
The most recent publication from the ATLAS Collaboration \cite{Aaboud:2017rwm} includes jet variables as well.
The decay modes of the \PZ boson to electron and muon ($\ell = \Pe,\mu$) pairs have been exploited.
Reconstructed distributions are corrected for event selection efficiency and detector resolution effects by means of an iterative unfolding technique, which
makes use of a response matrix to map physics variables at generator level onto their reconstructed values.

This Letter presents the dependence of the cross section on the jet multiplicity and the kinematic properties of the two \pt-leading jets (where \pt is the transverse momentum).
Comparison with theoretical predictions provides an important test of the QCD corrections to $\PZ\PZ$ production. Normalized differential cross sections as a function of the \pt and
pseudorapidity $\eta$ of the two \pt-leading jets, as well as their invariant mass (\mjj) and pseudorapidity separation (\detajj), are presented.
The study of \mjj establishes the basis for future multiboson final-state searches and for the investigation of phenomena involving interactions with four bosons at a single vertex,
while the measurement of the \detajj distribution is instrumental in the study of vector boson scattering.
The analysis presented in this paper together with the analyses reported in~\cite{ZZXS7TeVPaper,ZZXS8TeVPaper, ZZXS13TeV2015Paper, ZZXS13TeVPaper,ZZVBS13} seeks a detailed understanding of the SM processes that generate four leptons in the final state through the production of two Z bosons.
All measurements are compared to predictions from recent Monte Carlo (MC) event generators.
The data sets correspond to integrated luminosities of 19.7 and 35.9\fbinv, collected by the CMS Collaboration at 8 and 13\TeV, respectively.

\section{The CMS detector}
\label{sec:cmsandselection}

The central feature of the CMS apparatus is a superconducting solenoid of 6\unit{m} internal diameter, providing a magnetic field of 3.8\unit{T}.
Within the solenoid volume are silicon pixel and strip tracking detectors, a lead tungstate crystal electromagnetic calorimeter (ECAL), and a brass and scintillator
hadron calorimeter (HCAL), each composed of a barrel and two endcap sections. Forward calorimeters extend the $\eta$ coverage provided by the barrel and endcap detectors
up to $\abs{\eta} = 5$. Muons are measured in gas-ionization detectors embedded in the steel flux-return yoke outside the solenoid, using three different technologies:
drift tubes for $\abs{\eta} < 1.2$, cathode strip chambers for $0.9 < \abs{\eta} < 2.4$, and resistive plate chambers for $\abs{\eta} < 1.6$.
The silicon tracker measures charged particles within the range $\abs{\eta}< 2.5$. For nonisolated particles in the range $1 < \pt < 10\GeV$ and $\abs{\eta} < 1.4$,
the track resolutions are typically 1.5\% in \pt and 25--90 (45--150)\mum in the transverse (longitudinal) impact parameter \cite{TRK-11-001}.

The first level of the CMS trigger system~\cite{Khachatryan:2016bia}, composed of custom hardware processors, uses information from the calorimeters and muon detectors to select the most interesting events within a time interval of less than 4\mus. The high-level trigger processor farm further decreases the event rate from around 100\unit{kHz} to less than 1\unit{kHz}, before data storage.

A more detailed description of the CMS detector, together with a definition of the coordinate system used and the relevant kinematic variables, can be found in Ref.~\cite{Chatrchyan:2008zzk}.

\section{Signal and background simulation}
\label{sec:mc}

Several MC event generators are used to simulate the signal and background contributions.
The MC simulation samples are employed to optimize the event selection, evaluate the signal efficiency and acceptance,
estimate part of the background, and extract the unfolding response matrices used to correct for detector effects in the measured distributions.

For the 8\TeV data analysis, \MADGRAPHf~1.3.3~\cite{Alwall:2007fs,MGatNLO} is used to simulate the production of the
four-lepton final state at leading order (LO) in QCD with up to 2 jets included in the matrix-element calculations.
\POWHEG~2.0~\cite{Melia:2011tj, Nason:2004rx,PowhegMethod,Alioli:2010xd}
is used for the simulation of the same process at next-to-leading-order (NLO).
A sample of events generated with \MGvATNLO 2.3.3 (abbreviated as \MGNLO in the following)~\cite{Frederix:2012ps,MGatNLO}, which simulates signal processes at NLO with zero and one jet included in the matrix-element calculations, is produced only at generator level and used for comparison purposes.
For the 13\TeV data analysis, the four-lepton processes are simulated at NLO in QCD with 0 or 1 jet included in the matrix-element calculations with \MGNLO and with \POWHEG~2.0 at NLO.
The latter is scaled by a factor of 1.1 to reproduce the total $\PZ\PZ$ production cross section calculated at next-to-next-to-leading order (NNLO)~\cite{Cascioli:2014yka} at 13\TeV.
\MGNLO and \POWHEG~2.0, for both the 8 and 13\TeV analyses, include $\PZ\PZ$, $\PZ\gamma^\ast$, $\PZ$, and $\gamma^\ast\gamma^\ast$
processes, with the generator level constraint $m_{\ell^+\ell^{-}} > 4\GeV$ applied to all pairs
of oppositely charged same-flavor leptons, to avoid infrared divergences.

The $\cPg\cPg \to \PZ\PZ$ processes, which occur via loop-induced diagrams, are generated at LO with \MCFM~6.7 (7.0)~\cite{MCFM} for the 8 (13)\TeV analysis.
The 13\TeV samples are scaled by a factor of 1.7 to match the cross section computed at NLO~\cite{Caola:2015psa}. Electroweak production of four leptons and two jets
is simulated at LO with \PHANTOM~\cite{Phantom}. This sample includes triboson processes, where
the \PZ boson pair is accompanied by a third vector boson that decays into jets, as well as diagrams with quartic vertices.

Other diboson and triboson processes ($\PW\PZ$, $\PZ\gamma$, $\PW\PW\PZ$) as well as $\ttbar\PZ$, \ttbar, and \zjets samples are
generated at LO with \MADGRAPHf for the 8\TeV analysis, and at NLO with \MGNLO, for the 13\TeV analysis.

For the 8\TeV analysis, the \PYTHIA~6.4.24~\cite{Sjostrand:2006za} package, with parameters set by the Z2* tune~\cite{Chatrchyan:2013gfi}, is used for parton showering, hadronization,
and the underlying event simulation for all MC samples except for \MGNLO, for which  \PYTHIA~8.205~\cite{Sjostrand:2015}
is employed.
The default sets of parton distribution functions (PDFs) are CTEQ6L~\cite{CTEQ6L}
for the LO generators, and CT10~\cite{CT10}, for the NLO ones.
For the 13\TeV analysis, \PYTHIA~8.212~\cite{Sjostrand:2015}, with parameters set by the CUETP8M1 tune~\cite{Khachatryan:2015pea}, is
used for parton showering, hadronization, and the underlying event simulation.
The NNPDF3.0~\cite{NNPDF2015} PDF set is the default.
For all simulated event samples, the PDFs used are evaluated at the same order in QCD as the process in the sample.

The detector response is simulated using a detailed
description of the CMS detector implemented with the \GEANTfour
package~\cite{GEANT}. The simulated events are reconstructed with
the same algorithms used for the data.
The simulated samples include additional interactions per bunch crossing,
referred to as pileup.
Simulated events are weighted so that the pileup distribution reproduces that observed in the
data, with an average of about 21 (27) interactions per bunch
crossing for the 8 (13)\TeV data set.

\section{Particle reconstruction and event selection}
\label{sec:eventselection}

The primary triggers for this analysis require the presence of two loosely isolated leptons of the same or of different flavor.
The minimum \pt for the first lepton is 17\GeV, while it is 8 (12)\GeV for the second lepton in the 8 (13)\TeV analysis.
Triggers requiring a triplet of low-\pt leptons with no isolation requirement and, for the 13\TeV analysis, isolated single-electron and single-muon triggers, with minimal
\pt-thresholds of 27 and 22\GeV, respectively, help to increase the efficiency. The overall trigger efficiency for events that pass the
$\PZ\PZ$ selection is greater than 98\%.

The offline event selection procedure is similar to that of the inclusive $\PZ\PZ$ analyses~\cite{ZZXS8TeVPaper, ZZXS13TeV2015Paper,ZZXS13TeVPaper}
and is based on a global event description~\cite{CMS-PRF-14-001} that classifies particles into mutually exclusive categories:
charged hadrons, neutral hadrons, photons, muons, and electrons.
Events are required to have at least one vertex~\cite{TRK-11-001} within 24\unit{cm} of the geometric center of the
detector along the beam direction, and within 2\unit{cm} in the transverse plane.
Because of pileup the selected event can have several reconstructed vertices.

For the analysis at 8\TeV the vertex with the largest sum of the $\pt^2$ of the tracks
associated to it is chosen as the primary $\Pp\Pp$ interaction vertex, while at 13\TeV the reconstructed vertex with the largest value of summed
physics-object $\pt^2$ is taken to be the primary vertex.
The physics objects are the objects returned by a jet finding algorithm~\cite{Cacciari:2008gp,Cacciari:2011ma} applied to all charged tracks associated with the vertex, and the associated missing \pt, taken as the negative vector sum of the \pt of those jets.
Events with leptons are selected by requiring each lepton track to have a transverse impact parameter, with respect to
the primary vertex, smaller than 0.5\unit{cm} and a longitudinal impact parameter smaller than 1.0\unit{cm}.

Electrons are measured in the range $\abs{\eta} < 2.5$ by using both the tracking system and the ECAL. They are identified by means of a multivariate
discriminant that includes observables sensitive to
bremsstrahlung along the electron trajectory, the geometrical and momentum-energy agreement between the
electron track and the associated energy cluster in the ECAL, the shape of the electromagnetic shower,
and variables that discriminate against electrons originating from photon conversions~\cite{Khachatryan:2015hwa}.
The momentum resolution for electrons with $\pt \approx 45\GeV$ from $\PZ \to \Pe^+ \Pe^-$ decays ranges from 1.7\% for nonshowering electrons in the
barrel region to 4.5\% for showering electrons in the endcaps~\cite{Khachatryan:2015hwa}.

Muons are reconstructed in the range $\abs{\eta} < 2.4$ by combining information from the silicon tracker and the muon system~\cite{Chatrchyan:2012xi}.
The matching between the inner and outer tracks proceeds either outside-in, starting from a track in the muon system,
or inside-out, starting from a track in the silicon tracker.
The muons are selected among the reconstructed muon track candidates by applying minimal requirements on the track in both
the muon system and the inner tracker system, and taking into account the compatibility with minimum-ionizing particle energy deposits in the calorimeters.
In the intermediate range of $20<\pt<100\GeV$, matching muons to tracks measured in the silicon tracker results in a relative \pt resolution of 1.3--2.0\% in the barrel,
and better than 6\% in the endcaps. The \pt resolution in the barrel is better than 10\% for muons with \pt up to 1\TeV~\cite{Chatrchyan:2012xi}.

Electrons (muons) are considered candidates for inclusion in the four-lepton final states if they have $\ptl>7\,(5)\GeV$ and $\aetal<2.5\,(2.4)$.
In order to suppress electrons from photon conversions and muons originating from in-flight decays of hadrons,
we place a requirement on the impact parameter computed in three dimensions. We require that the ratio of the impact parameter for the track and its uncertainty to be less than 4.
To discriminate between prompt leptons from \PZ boson decay and those arising from electroweak decays of hadrons within jets,
an isolation requirement for leptons is imposed. The relative isolation is defined as
\begin{equation}
        R_\text{iso} = \bigg[ \sum_{\substack{\text{charged} \\ \text{hadrons}}} \!\! \pt \, + \,
                             \max\big(0, \sum_{\substack{\text{neutral} \\ \text{hadrons}}} \!\! \pt
                                       \, + \, \sum_{\text{photons}} \!\! \pt \, - \, \pt^\mathrm{PU}
                                       \big)\bigg] \bigg/ \ptl,
        \label{eq:iso}
\end{equation}
where the sums run over the charged and neutral hadrons, and photons, in a
cone defined by
$\DR \equiv \sqrt{\smash[b]{(\Delta\eta)^2+(\Delta\phi)^2}}$ around the lepton trajectory. The radius \DR{} is set to be 0.4 and 0.3 in the
8 and 13\TeV data analyses, respectively.
To minimize the contribution of charged particles from pileup to the isolation calculation,
charged hadrons are included only if they originate from the
primary vertex. The contributions of neutral particles from pileup to the activity inside the cone around a lepton is referred to as
$\pt^\mathrm{PU}$, and is obtained with different methods for electrons and muons.
For electrons, $\pt^\mathrm{PU}$
is evaluated with the jet area  method described in Ref.~\cite{Cacciari:2007fd}.
For muons, it is taken
to be half the sum of the \pt of all charged particles in the cone originating
from pileup vertices.
The factor of one-half accounts for the expected fraction of neutral to charged particles in hadronic interactions.
A lepton is considered isolated if
$R_\text{iso} < 0.4\,(0.35)$ in the 8 (13)\TeV data analysis.

The lepton momentum scales are calibrated in bins of \ptl and $\eta_\ell$
using the decay products of known resonances decaying to lepton pairs. The measured lepton momentum
scale is corrected with a $\PZ\to \ell^+\ell^-$ sample, by
matching the peak of the reconstructed dilepton mass spectrum to the nominal
value of $m_{\PZ}$~\cite{Zmass}. Muon momenta are calibrated by using \JPsi decays as well.
We account for final-state radiation of leptons by correcting their momenta with photons of $\pt > 2\GeV$
and within a cone of $\DR = 0.5$ around the lepton momentum direction~\cite{Chatrchyan:2013mxa,Sirunyan:2017exp}.
The photons selected by this algorithm are excluded from the lepton isolation computation.
The efficiency of the lepton reconstruction and selection is
measured with the tag--and--probe technique~\cite{CMS:2011aa} in bins of \ptl
and $\eta_\ell$. This measurement is used to correct the simulation efficiency.

Jets are reconstructed from particle candidates by means of the anti-\kt clustering algorithm~\cite{Cacciari:2008gp}, as implemented in
the \FASTJET package~\cite{Cacciari:2011ma}, with a distance parameter of 0.5 (0.4) in the 8~(13)\TeV data analysis.
The jet energy resolution amounts typically to 15\% at 10\GeV, 8\% at 100\GeV, and 4\% at 1\TeV.

Jet energy corrections are extracted from the data and the simulated events by combining several measurements and methods that
account for the effects of pileup, non-uniform detector response, and residual data-simulation jet energy scale (JES) differences.
The JES  calibration~\cite{JESCalib, Khachatryan:2016kdb} relies on corrections parametrized in terms of the uncorrected
\pt and $\eta$ of the jet, and are applied as multiplicative factors to the four-momentum vector of each jet.

In order to maximize the reconstruction efficiency while reducing the
instrumental background and contamination from pileup jets, loose
identification quality criteria~\cite{CMS-PAS-JME-16-003} are imposed on jets, based on the
energy fraction carried by charged and neutral hadrons, as well as
charged leptons and photons.
A minimum threshold of 30\GeV on the \pt of jets is required to ensure that they are well measured and to reduce the pileup contamination.
Jets are required to have $\abs{\eta} < 4.7$ and to be separated from all selected lepton candidates by at least
$\DR = 0.5\,(0.4)$ in the 8 (13)\TeV analysis.

A signal event must contain at least two
$\PZ/\gamma^{\ast}$ candidates, each reconstructed from a pair of isolated electrons or muons of opposite charges.
The highest-\pt lepton must have $\pt > 20\GeV$, and
the second-highest lepton $\pt^\Pe > 10\,(12)\GeV$ if it is an electron,
or $\pt^\Pgm > 10\GeV$ in case of a muon for the analysis at $\sqrt{s}=8\,(13)\TeV$.
All leptons are required to be separated by
$\DR \left(\ell, \ell' \right) > 0.02$, and electrons are required to be separated from muons by
$\DR \left(\Pe, \mu \right) > 0.05$.

Within each event, all permutations of oppositely charged leptons
giving a valid pair of $\PZ/\gamma^{\ast}$
candidates are considered separately.
For each $4\ell$ candidate, the lepton pair with the invariant mass closest to the nominal \PZ boson mass is denoted by $\PZ_1$
and the other dilepton candidate is denoted by $\PZ_2$.
Both $\PZ_1$ and $\PZ_2$ are required to have a mass between 60 and 120\GeV.
All pairs of oppositely charged leptons in the $4\ell$ candidate are required to
have $m_{\ell \ell'} > 4\GeV$ regardless of their flavor to remove contributions from the decay of low-mass hadron resonances.

If multiple $4\ell$
candidates within an event pass this selection, the candidate with $m_{\PZ_1}$ closest to
the nominal \PZ boson mass is chosen. In the rare cases (0.3\%) of further ambiguity, which may arise in events with
 more than 4 leptons, the $\PZ_2$ candidate that maximizes the scalar \pt sum of the four leptons is chosen.
The set of selection criteria just described is referred to as the $\PZ\PZ$ selection, and gives a total of 288~(927)
observed events at $\sqrt{s}=8\,(13)\TeV$. The corresponding number of expected signal events from MC prediction is about 271~(850).

\section{Background estimation}
\label{sec:bckgestimation}

The largest source of background arises from processes in which heavy-flavor jets produce
secondary leptons, and from processes in which jets are misidentified as leptons.
The main contributing processes are \zjets, \ttbar, and \wzjets.

However, the lepton identification and isolation requirements reduce
this background to a very small level compared to the signal. The
residual contribution is estimated from data samples consisting of \PZ+$\ell\ell$
events that are required to pass the $\PZ\PZ$ selection described in
Section~\ref{sec:eventselection}, except that either one or both leptons belonging to the $\PZ_2$
candidate fail the isolation or identification requirements. Two
control samples are selected, with one and two misidentified leptons,
respectively. The background yield in the signal region is estimated by
weighting the number of events in the control samples by the lepton
misidentification rate measured in data in a dedicated control region.
The procedure is identical to that of Refs.~\cite{ZZXS13TeV2015Paper,ZZXS13TeVPaper}
and is described in more detail in Ref.~\cite{Chatrchyan:2013mxa}.

Another source of background arises from processes that produce four genuine
high-\pt isolated leptons, $\Pp\Pp \to \ttbar\PZ$ and $\Pp\Pp \to \PW\PW\PZ$.
This contribution is small and is estimated by using the corresponding simulated samples.

The total estimated background yields are $8 \pm 4$ ($37 \pm 11$) events in the 8 (13)\TeV signal region.

\section{Systematic uncertainties}
\label{sec:syst}

The systematic uncertainties are estimated by varying the quantities that may affect the cross section and by propagating the changes to the analysis procedure.
The systematic uncertainties from sources that may affect the differential cross section shapes have been estimated through the unfolding procedure by recomputing the response matrix, after varying each source of systematic uncertainty independently and in both directions, up and down.
The systematic uncertainties in the differential cross section as a function of the jet multiplicity are summarized in Table~\ref{tab:syst}. Those that depend on the number of jets in the event are listed as a range.

The systematic uncertainty in the trigger efficiency is evaluated by taking the difference between the value obtained from the data 
and that from the simulated events, and it leads to a 1.5 (2.0)\% uncertainty in the differential cross sections measured with the 8 (13)\TeV data. 
The uncertainties arising from lepton reconstruction and selection (identification, isolation, and impact parameter determination)
depend on the jet multiplicity, are sensitive to statistical fluctuations, and range between 0.9 and 4.4\%, in the 8\TeV analysis (3.7 and 4.5\%, in the 13\TeV analysis).
The largest contribution to the systematic uncertainty in the differential cross section measurements comes from the JES determination, which increases with the jet multiplicity and reaches 9.2 (17.5)\% when the number of jets exceeds two in the 8 (13)\TeV analysis.
Likewise, the uncertainty due to the jet energy resolution (JER) increases from 0.2 to 1.7\% (2.1 to 8.4\%) for the 8~(13)\TeV samples.
The larger JES and JER uncertainties for the 13 TeV sample reflect the increase in the number of soft jets (with \pt close to the 30\GeV threshold) as a function of
the center-of-mass energy.

The uncertainties in the \zjets, \wzjets, and \ttbar background have two components, which are added in quadrature. The first relates to the different relative fraction of these background processes in the control sample where we measure the lepton misidentification rate and the sample to which this rate is applied. The second is the statistical uncertainty in the control sample.
The effect of these uncertainties increases with the jet multiplicity and amounts to 0.7--6.9\% (0.5--2.4\%) in the 8~(13)\TeV measurement.
The contribution to the uncertainty from the modeling of genuine four lepton background is smaller and varies between 0.1 and 2.0\% (${<}0.1$ and 1.2\%) for the 8 (13)\TeV data.
The pileup uncertainty is evaluated by varying the pileup modeling in the MC samples within its uncertainty.
The uncertainty in the integrated luminosity is 2.6~\cite{CMS-PAS-LUM-13-001} and 2.5\%~\cite{CMS-PAS-LUM-17-001} for the 8 and 13\TeV data, respectively.

The contribution of the MC generator choice to the systematic uncertainty is obtained by  comparing the results found with two different sets of MC samples:
\MADGRAPHf + \MCFM + \PHANTOM (\MGNLO + \MCFM + \PHANTOM) and \POWHEG + \MCFM + \PHANTOM for the 8 (13)\TeV measurement,
and ranges from 0.2 to 3.7\% (0.5 to 5.0\%) at 8 (13)\TeV.
The impact of the relative contribution of the $\cPq\cPaq \to \PZ\PZ$ and $\cPg\cPg \to \PZ\PZ$ processes in the response matrix definition is less than 1\% and is evaluated by varying the corresponding cross section within their renormalization and factorization scale uncertainties. For 8\TeV, where no LO to NLO factor is applied to the \MCFM cross section,
the $\cPg\cPg \to \PZ\PZ$ cross section is varied by 100\% of its value.
The statistical uncertainties of the MC samples result in negligible contributions to the response matrix uncertainty.
The systematic uncertainty arising from the choice of the PDF and the strong coupling strength \alpS
has been evaluated using the PDF4LHC recommendations~\cite{PDF4LHCRec, PDF4LHCRep, PDFLHC}, using the CT10, MSTW08, and NNPDF2.3~\cite{NNPDF2011} PDF sets, in the 8\TeV analysis,
and the NNPDF3.0 set in the 13\TeV analysis.

The total systematic uncertainty is obtained by summing all the sources in quadrature, taking into account the correlations among the different channels.

For the normalized differential cross sections, only systematic uncertainties affecting the shape of the distributions are relevant.
The uncertainties in the luminosity and trigger efficiency cancel out completely, as well as other contributions to the uncertainty in the total yield.

\begin{table*}[!htb]
\centering
\topcaption{
The contributions to the uncertainty in the absolute and normalized differential cross section measurements in Fig.~\ref{fig:diff_xs_nJets_notnorm} and~\ref{fig:diff_xs_nJets}, upper panels. Uncertainties that depend on jet multiplicity are listed as a range.}
\label{tab:syst}
\cmsTable{
\begin{tabular}{lcccc}
\hline
                                   & \multicolumn{2}{c}{8\TeV data}       & \multicolumn{2}{c}{13\TeV data}     \\
Systematic source                   &  Absolute (\%)      &       Normalized (\%)                      & Absolute (\%)      & Normalized (\%)    \\
\hline
Trigger                             &        1.5   &                   \NA                         &       2.0   &     \NA     \\
Lepton reconstruction and selection &   0.9--4.4   &              ${\leq} 0.1$                     &  3.7--4.5   &  0.1--0.8   \\
Jet energy scale                    &   1.5--9.2   &               1.5--9.1                        & 4.6--17.5   &  4.6--17.5  \\
Jet energy resolution               &   0.2--1.7   &               0.2--1.7                        &  2.1--8.4   &  2.1--8.4   \\
Background yields                   &   0.7--7.2   &               0.7--5.4                        &  0.5--2.8   &  0.4--2.0   \\
Pileup                              &        1.8   &                    1.8                        &  0.3--1.9   &  0.6--1.8   \\
Luminosity                          &        2.6   &                   \NA                         &       2.5   &     \NA     \\
Choice of Monte Carlo generators    &   0.2--3.7   &               0.2--3.7                        &  0.5--5.0   &  0.8--4.7   \\
qq/gg cross section                 &   0.1--0.8   &               0.1--0.8                        & ${<}0.1$--0.3 &  0.1--0.2   \\
PDF                                 &        1.0   &                   \NA                         & ${<}0.1$--0.2 &  ${<}0.1$--0.2\\
\alpS                               &      ${<} 0.1$  &                 ${<}0.1$                   &${\leq} 0.1$    & ${\leq}0.1$   \\
\hline
\end{tabular}
}
\end{table*}

\section{The \zzjets differential cross section measurements}
\label{sec:xs}

\label{sec:fid_xsec}

\begin{figure*}[!hbtp]
  \centering
  \includegraphics[width=1.01\cmsFigWidth]{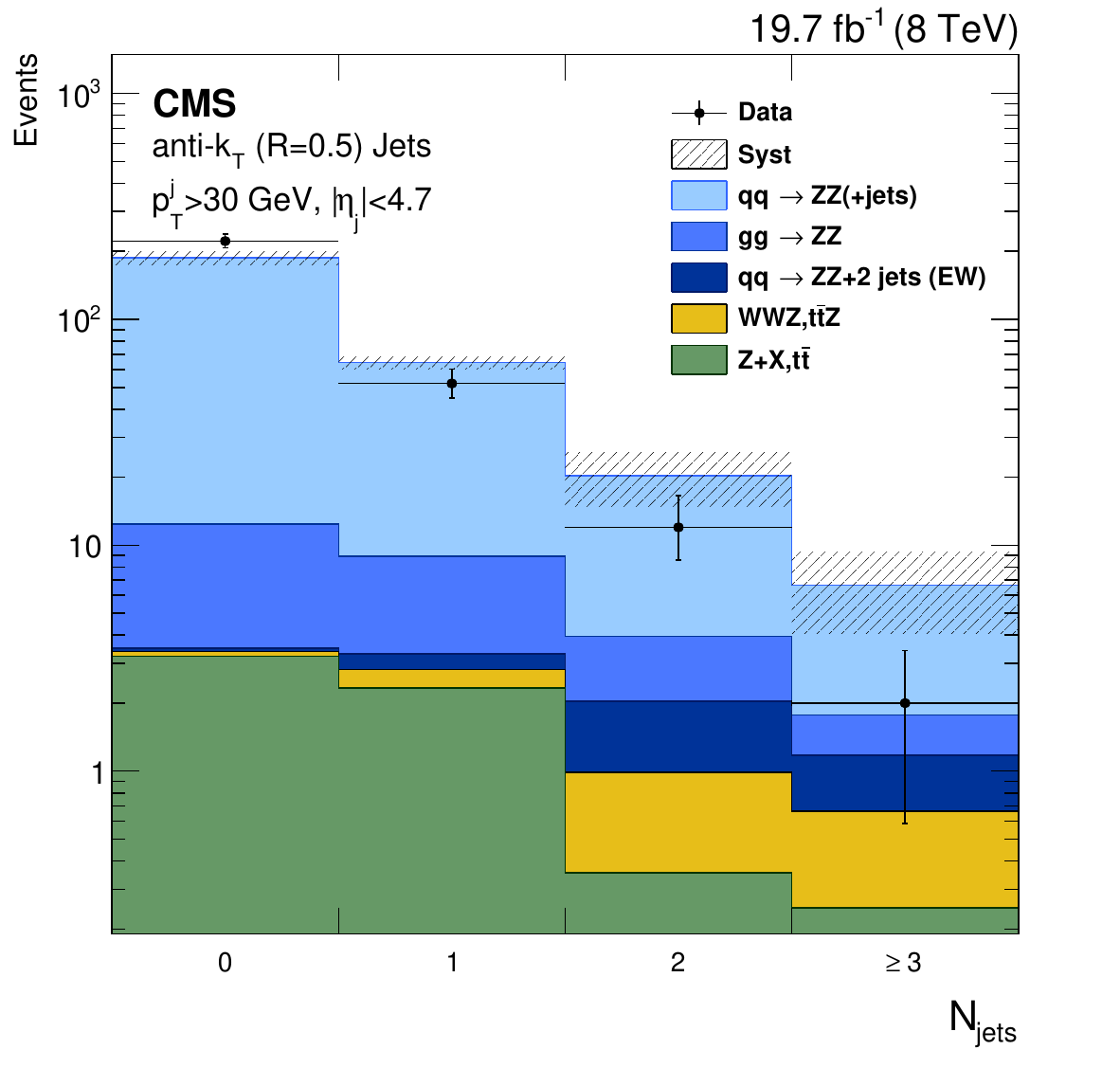}
    \includegraphics[width=1.01\cmsFigWidth]{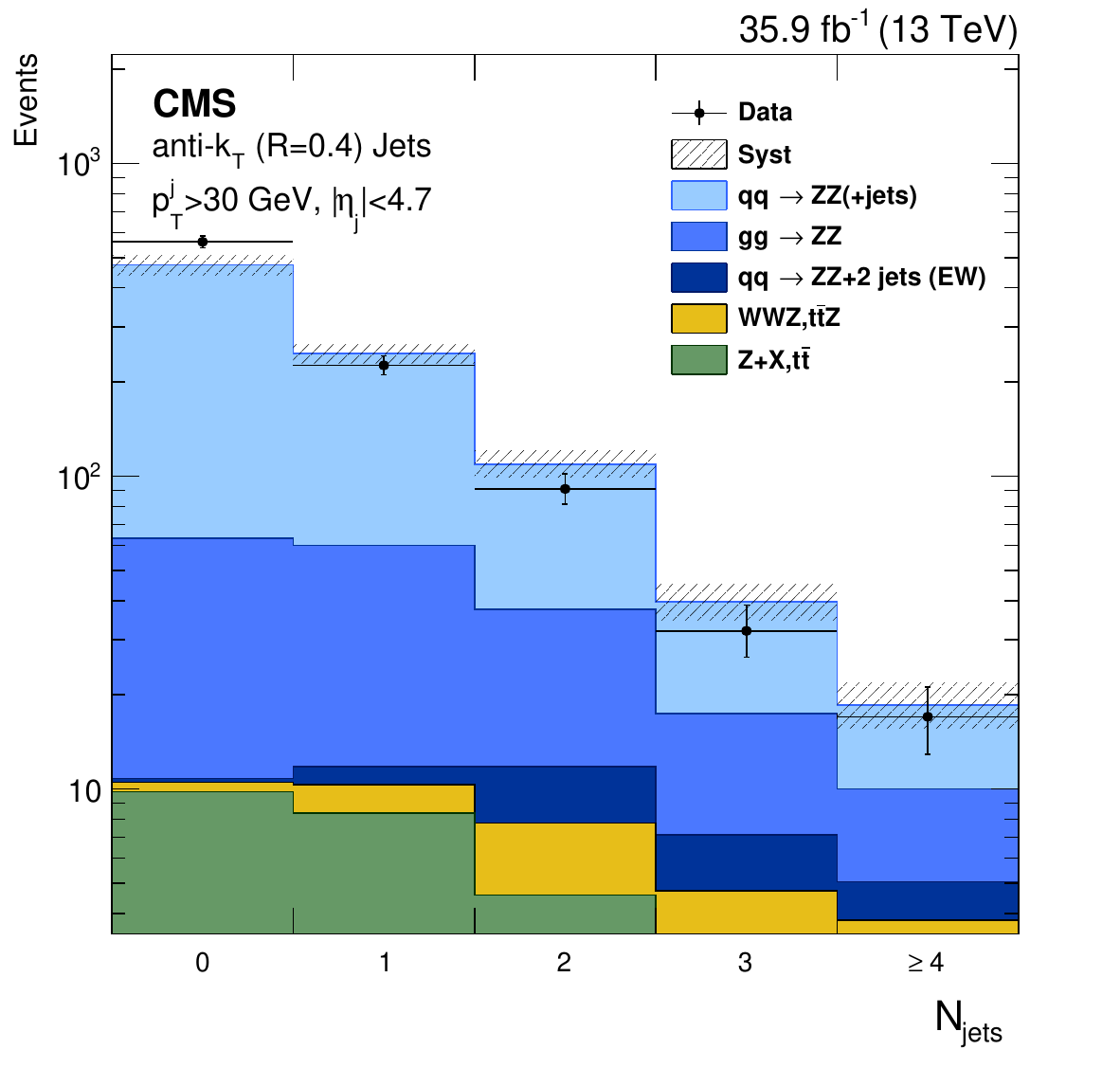}
   \caption{
Distribution of the reconstructed jet multiplicity in the 8\TeV (left) and 13\TeV (right) data. The points represent the data and the vertical bars correspond to the statistical uncertainty. The shaded histograms represent MC predictions and the background estimates, while the hatched band on their sum indicates the systematic uncertainty of the prediction. The \zjets and \ttbar background is obtained from the data.}
   \label{fig:njets_reco}
\end{figure*}

The distributions of the jet multiplicity combining the 4$\mu$, 4$\Pe$, and 2$\mu$2$\Pe$ channels are shown in Fig.~\ref{fig:njets_reco}, together with the SM expectations, the estimated backgrounds, and the systematic uncertainty in the prediction.

The differential $\Pp\Pp \to \PZ\PZ\to \ell\ell\ell'\ell'$ cross section is measured as a function of the jet multiplicity,
the \pt-leading jet transverse momentum ($\pt^{\mathrm{j1}}$) and pseudorapidity ($\eta_{\mathrm{j1}}$) with the 8 and 13\TeV data.
Because of the limited number of events with more than one jet at 8\TeV, the differential cross section as a function of the \pt-subleading jet transverse momentum ($\pt^{\mathrm{j2}}$)
and pseudorapidity ($\eta_{\mathrm{j2}}$), as well as the invariant mass of the two \pt-leading jets (\mjj) and their pseudorapidity separation (\detajj) are studied at 13\TeV only.
For all measurements we consider jets with $\pt^{\mathrm{j}} > 30\GeV$ and $\aetaj<4.7$. For the jet multiplicity distribution we also present the measurements made
with central jets ($\aetaj<2.4$) only. The measurements are performed for the two slightly different phase space regions
adopted for the 8~\cite{ZZXS8TeVPaper} and 13~\cite{ZZXS13TeVPaper}\TeV data, which are given in Table~\ref{tab:fiducial_ps}.
The generator-level lepton momenta are corrected by adding the momenta of generator-level photons within
$\DR\left(\ell,\gamma\right) < 0.1$. The \PZ bosons are then selected with the same method adopted to extract the signal at the reconstruction level.
In order to define the jets at generator level, the generated particles are clustered using the anti-\kt algorithm, with a distance parameter identical to the corresponding one at reconstruction level.

\begin{table}[htbp]
\centering
\topcaption{
Phase space definitions for cross section measurements at 8\TeV~\cite{ZZXS8TeVPaper} and 13\TeV~\cite{ZZXS13TeVPaper}.
The common definitions apply to both measurements.
}
\begin{tabular}{ll}
\hline
\multicolumn{1}{c}{8\TeV} & \multicolumn{1}{c}{13\TeV} \\
\hline
$\pt^{\Pe} > 7\GeV$, $\abs{\eta^{\Pe}} < 2.5 \ $  & $\pt^{\Pe} > 5\GeV$, $\abs{\eta^{\Pe}} < 2.5$ \\
$\pt^{\mu} > 5\GeV$, $\abs{\eta^{\mu}} < 2.4 \ $  & $\pt^{\mu} > 5\GeV$, $\abs{\eta^{\mu}} < 2.5$ \\ [\cmsTabSkip]
\multicolumn{2}{c}{Common definitions} \\ [\cmsTabSkip]
\multicolumn{2}{l}{ $\pt^{\ell_1} > 20\GeV$ ,   $\pt^{\ell_2} > 10\GeV$ } \\

\multicolumn{2}{l}{ $m_{\ell^+ \ell^-} > 4\GeV$ (any opposite-sign same-flavor pair)  } \\

\multicolumn{2}{l}{ $60 < (m_{\PZ_1}, m_{\PZ_2}) < 120\GeV$  }\\

\hline
\end{tabular}
\label{tab:fiducial_ps}
\end{table}

Each distribution is corrected for the event selection efficiency and the detector resolution effects by means of a response matrix that translates
the physics variables at generator level into their reconstructed values.
The correction procedure is based on the iterative D'Agostini
unfolding method technique~\cite{DAgostini}, as implemented in the \textsc{RooUnfold} toolkit~\cite{RooUnfold}, and regularized by stopping after four iterations.
The robustness of the result is tested against the singular value decomposition (SVD)~\cite{SVD} alternative unfolding method.
For each measured distribution, a response matrix is evaluated using two different sets of generators: the first
one includes \MADGRAPHf ($\cPq\cPaq \to \PZ\PZ$), \MCFM ($\cPg\cPg \to \PZ\PZ$)
and \PHANTOM ($\cPq\cPaq \to \PZ\PZ$ + 2 jets) for the 8\TeV data set and \MGNLO ($\cPq\cPaq \to \PZ\PZ$),
\MCFM ($\cPg\cPg \to \PZ\PZ$) and \PHANTOM ($\cPq\cPaq \to \PZ\PZ$ + 2 jets) for the 13\TeV data set.
In the second one, the \POWHEG sample is instead used for the $\cPq\cPaq \to \PZ\PZ$ process in both the 8 and 13\TeV data analyses.
The former set, where the leading-order MC generator can simulate up to two jets at matrix-element level, is taken as the reference, while the latter is used for comparison and to estimate the systematic uncertainty due to the MC generator choice.
After the unfolding, the cross sections for $\Pp\Pp \to \PZ\PZ + N\,\text{jets}\to \ell\ell\ell'\ell' + N\,\text{jets}$,
for $N = 0$, 1, 2, and $\ge$3, are extracted.

\begin{figure*}[!hbtp]
  \centering
     \includegraphics[width=0.95\cmsFigWidth]{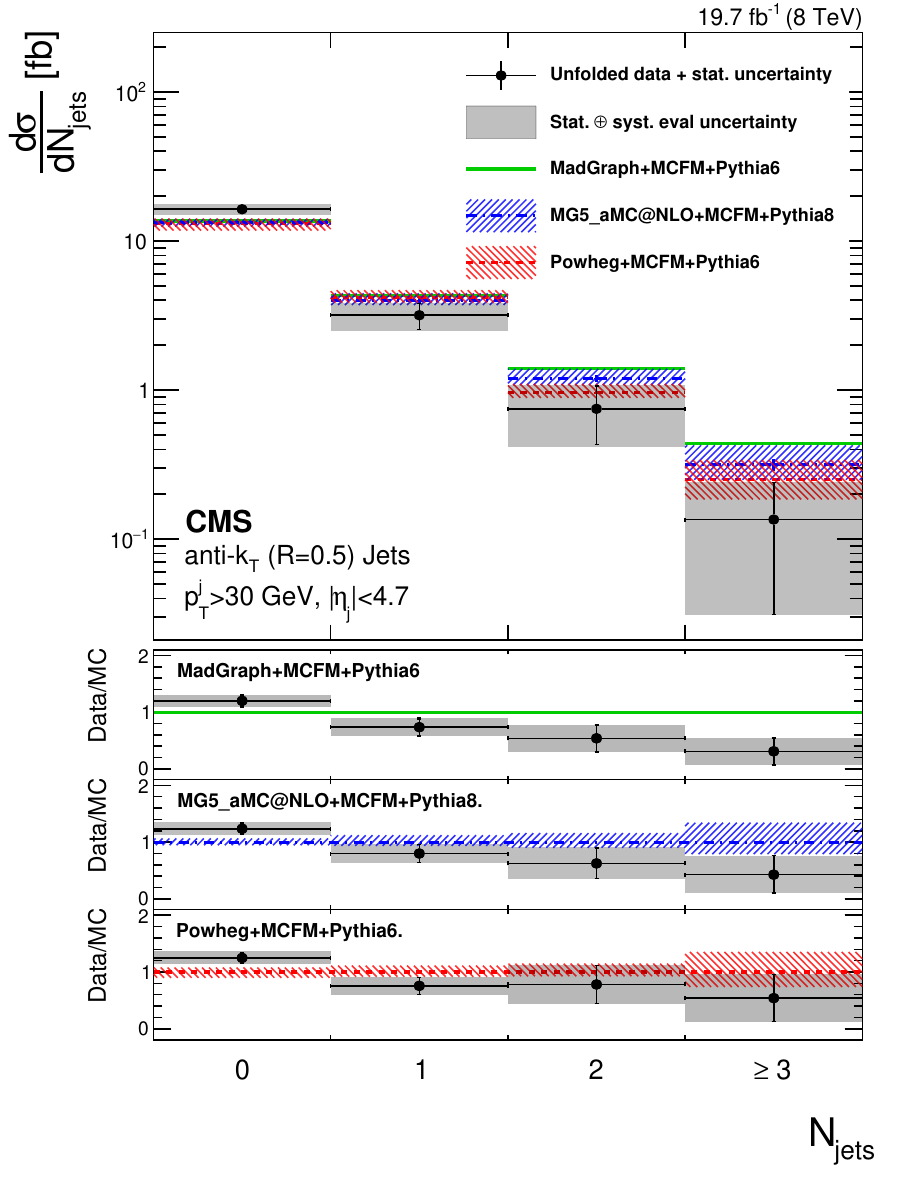}
     \includegraphics[width=0.95\cmsFigWidth]{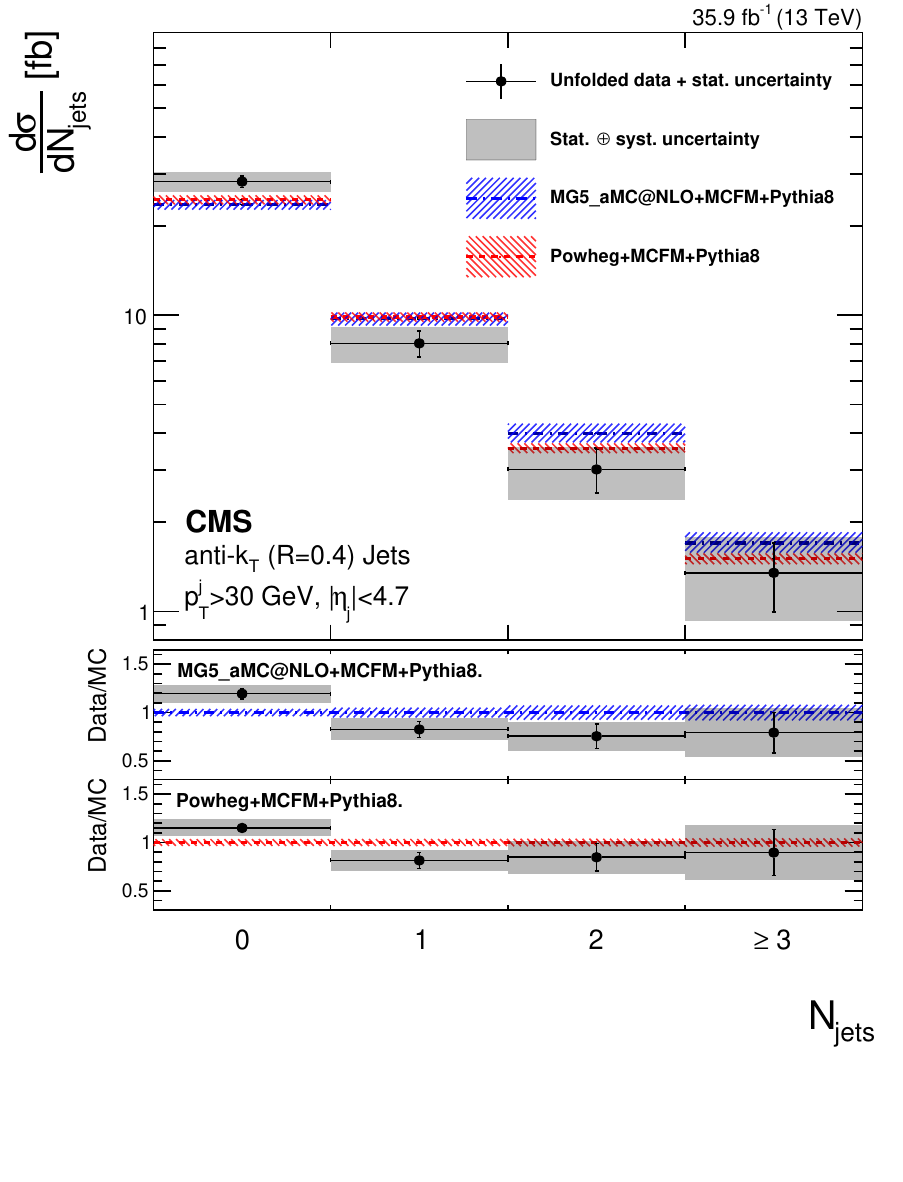}
     \includegraphics[width=0.95\cmsFigWidth]{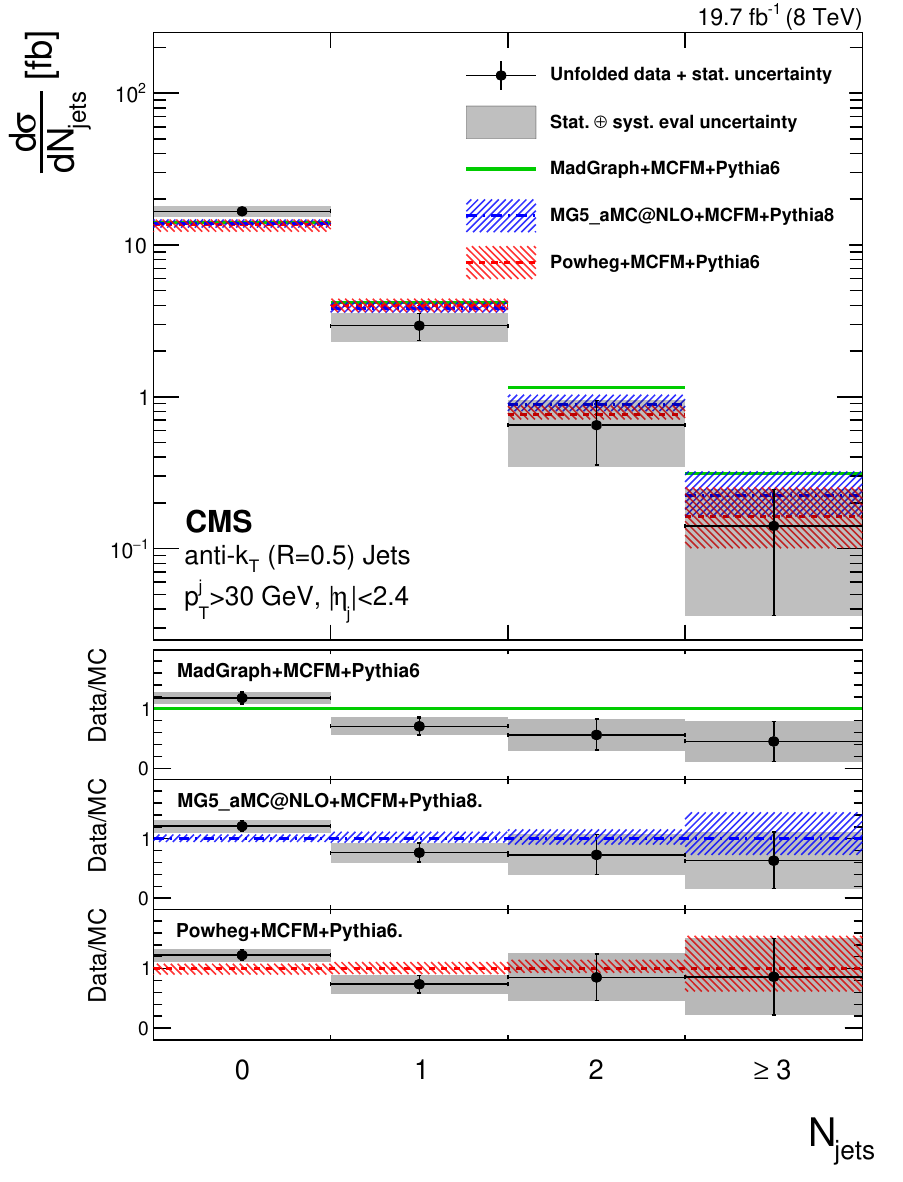}
     \includegraphics[width=0.95\cmsFigWidth]{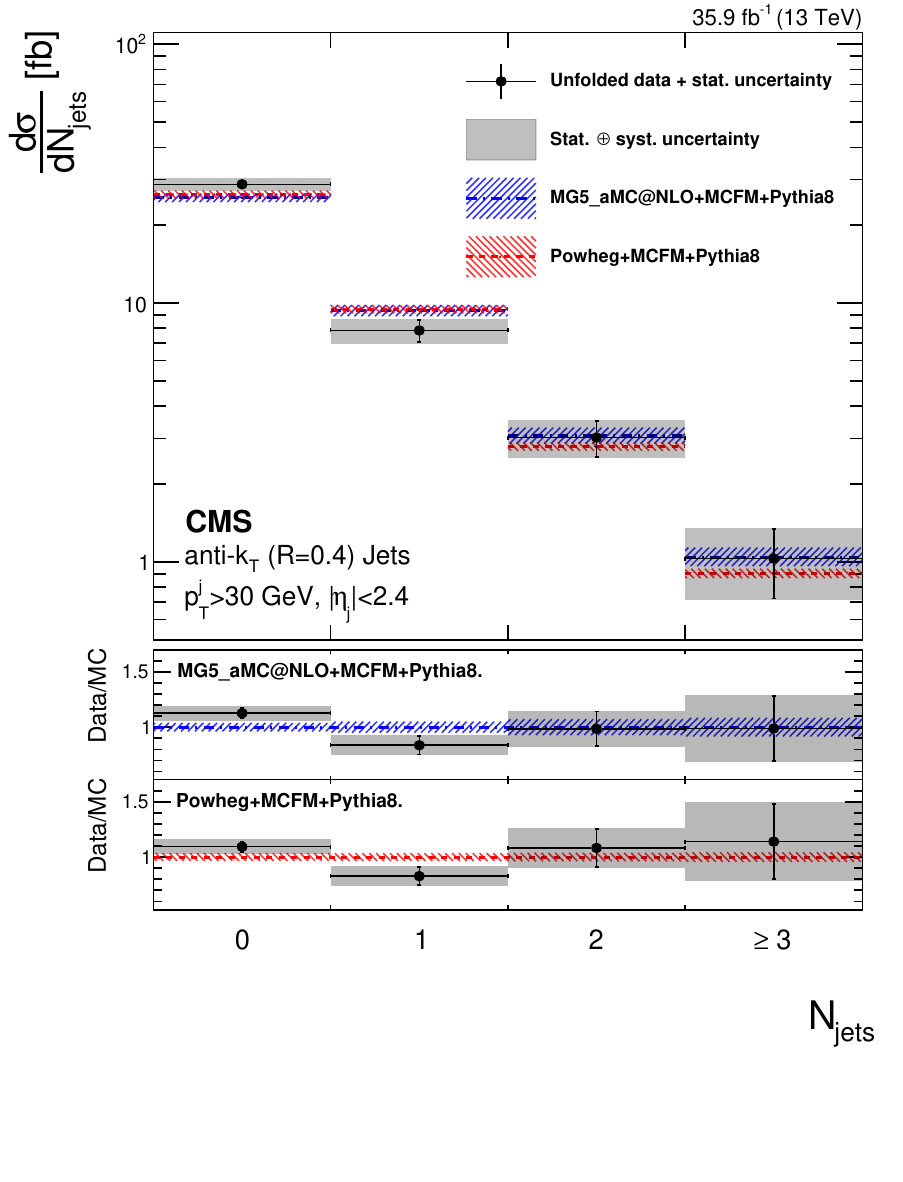}
   \caption{Differential cross sections of $\Pp\Pp\to\PZ\PZ\to 4\ell$ as a function of the multiplicity of jets with $\aetaj<4.7$  (top panels) and $\aetaj<2.4$ (bottom panels), for the 8 (left) and 13 (right)\TeV data.
The measurements are compared to the predictions of \MGNLO, \POWHEG, and \MADGRAPHf (8\TeV only) sets of samples.
Each MC set, along with the main MC generator, includes the \MCFM and \PHANTOM generators. \PYTHIA 6 and \PYTHIA 8 are used for parton showering, hadronization, and underlying event simulation, for the 8 and 13\TeV analysis, respectively, with the sole exception of \MGNLO, which is always interfaced to \PYTHIA 8.
The total experimental uncertainties are shown as hatched regions, while the colored bands display the theoretical uncertainties in the matrix-element calculations.
}
   \label{fig:diff_xs_nJets_notnorm}
\end{figure*}

The differential cross sections as a function of the jet multiplicity are shown in Fig.~\ref{fig:diff_xs_nJets_notnorm} for
$\aetaj < 4.7$ (upper) and for $\aetaj < 2.4$ (lower).
The ratios between the measured and expected distributions from the \MADGRAPHf, \MGNLO, and \POWHEG set of samples for $\sqrt{s}= 8\TeV$, and \POWHEG and \MGNLO for
$\sqrt{s}= 13\TeV$ are also shown in the figures.
Uncertainties in the MC predictions at the matrix-element level are evaluated by varying the renormalization and factorization scales independently, up and down, by a
factor of two with respect to the default values of $\mu_R = \mu_F = m_{4\ell}$ for \POWHEG and $\mu_\mathrm{R} = \mu_\mathrm{F} = \frac{1}{2} \sum \pt^{\mathrm{j}}$~+~$\sum \ptl$ for \MGNLO.
In the \MCFM predictions, the uncertainty in the LO to NLO cross section scaling factor includes the renormalization and factorization scales uncertainty.
The theoretical uncertainties also include  the uncertainties in the PDF and \alpS.
The measured and expected cross section values for $\aetaj < 4.7$ are given in Tables~\ref{tab:xs_8TeV_njets} and \ref{tab:xs_13TeV_njets}.

\begin{table*}[!htb]
\centering
\topcaption{
The $\Pp\Pp \to \PZ\PZ\to \ell\ell\ell'\ell'$ cross section at $\sqrt{s} = 8\TeV$ as a function of the jet multiplicity. The integrated luminosity uncertainty for number of jets = 2 and $\ge$3 is negligible and not quoted. The cross sections are compared to the theoretical predictions (last column) from \MGNLO+~\MCFM~+~\PHANTOM.}
\label{tab:xs_8TeV_njets}
\cmsTable{
\begin{tabular}{ccc}
\hline
Number of jets ($\aetaj < 4.7$) & Cross section [fb] & Theoretical cross section [fb] \\
\hline

0      & $16.3 \pm 1.2\stat^{+ 1.0 }_{- 0.9 }\syst \pm 0.4\lum$ & $ 13.2 ^{+ 0.9 }_{- 0.7 }$ \\
1      & $3.2  \pm 0.6\stat^{+ 0.3 }_{- 0.3 }\syst \pm 0.1\lum$ & $ 4.0  ^{+ 0.5 }_{- 0.3 }$ \\
2      & $0.7  \pm 0.3\stat^{+ 0.1 }_{- 0.1 }\syst $                      & $ 1.2 ^{+ 0.2 }_{- 0.1 }$ \\
$\ge$3 & $0.14 \pm 0.1\stat^{+ 0.01}_{-0.01 }\syst $                      & $ 0.3 ^{+ 0.1 }_{- 0.1 }$ \\
\hline
\end{tabular}
}
\end{table*}

\begin{table*}[!htb]
\centering
\topcaption{The $\Pp\Pp \to \PZ\PZ\to \ell\ell\ell'\ell'$ cross section at $\sqrt{s} = 13\TeV$ as a function of the jet multiplicity. The integrated luminosity uncertainty for the number of jets $\ge$3 is smaller than 0.1\unit{fb} and is not quoted. The cross sections are compared to the theoretical predictions (last column) from \MGNLO+~\MCFM~+~\PHANTOM.}
\label{tab:xs_13TeV_njets}
\cmsTable{
\begin{tabular}{ccc}
\hline
Number of jets ($\aetaj < 4.7$) & Cross section [fb] & Theoretical cross section [fb] \\
\hline
0      & $28.3 \pm 1.3\stat^{+ 1.7 }_{- 1.5 }\syst \pm 0.7\lum$ & $ 23.6 ^{+ 0.8 }_{- 0.9 }$\\
1      & $8.0  \pm 0.8\stat^{+ 0.7 }_{- 0.8 }\syst \pm 0.2\lum$ & $ 9.7 ^{+ 0.5 }_{- 0.5 }$\\
2      & $3.0  \pm 0.5\stat^{+ 0.3 }_{- 0.4 }\syst \pm 0.1\lum$ & $ 4.0 ^{+ 0.3 }_{- 0.3 }$\\
$\ge$3 & $1.3  \pm 0.4\stat^{+ 0.2 }_{- 0.2 }\syst$ & $ 1.7 ^{+ 0.1 }_{- 0.1 }$\\
\hline
 \end{tabular}
}
\end{table*}

\begin{figure*}[!hbtp]
  \centering
   \includegraphics[width=0.95\cmsFigWidth]{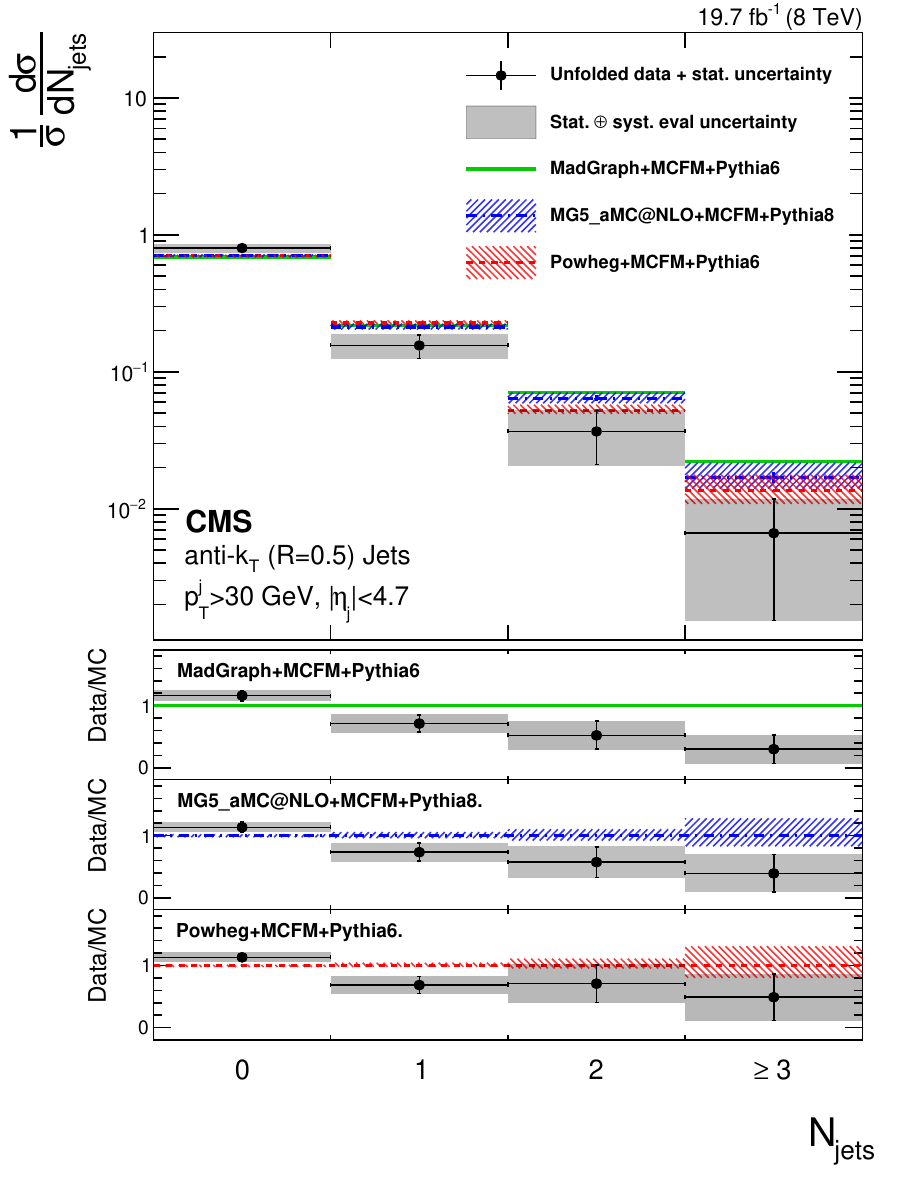}
   \includegraphics[width=0.95\cmsFigWidth]{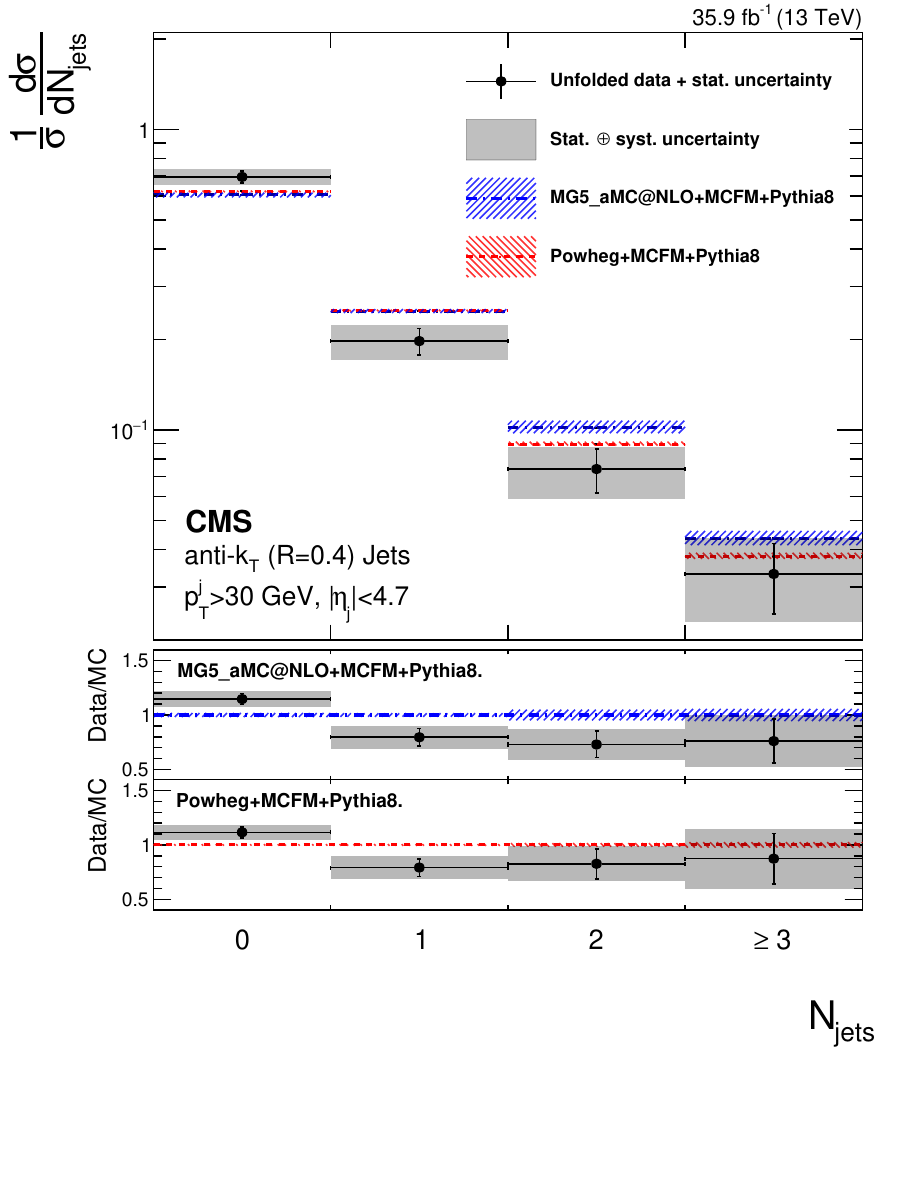}
   \includegraphics[width=0.95\cmsFigWidth]{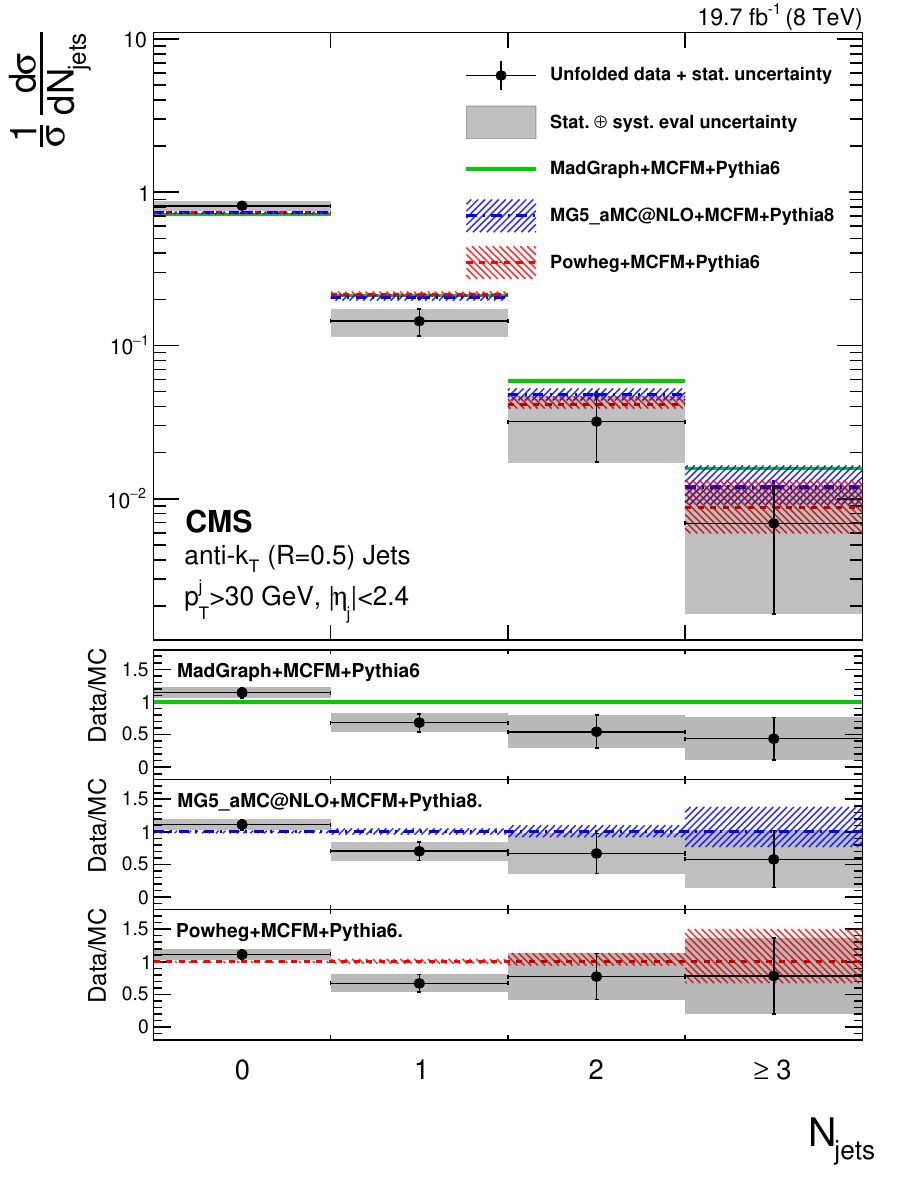}
   \includegraphics[width=0.95\cmsFigWidth]{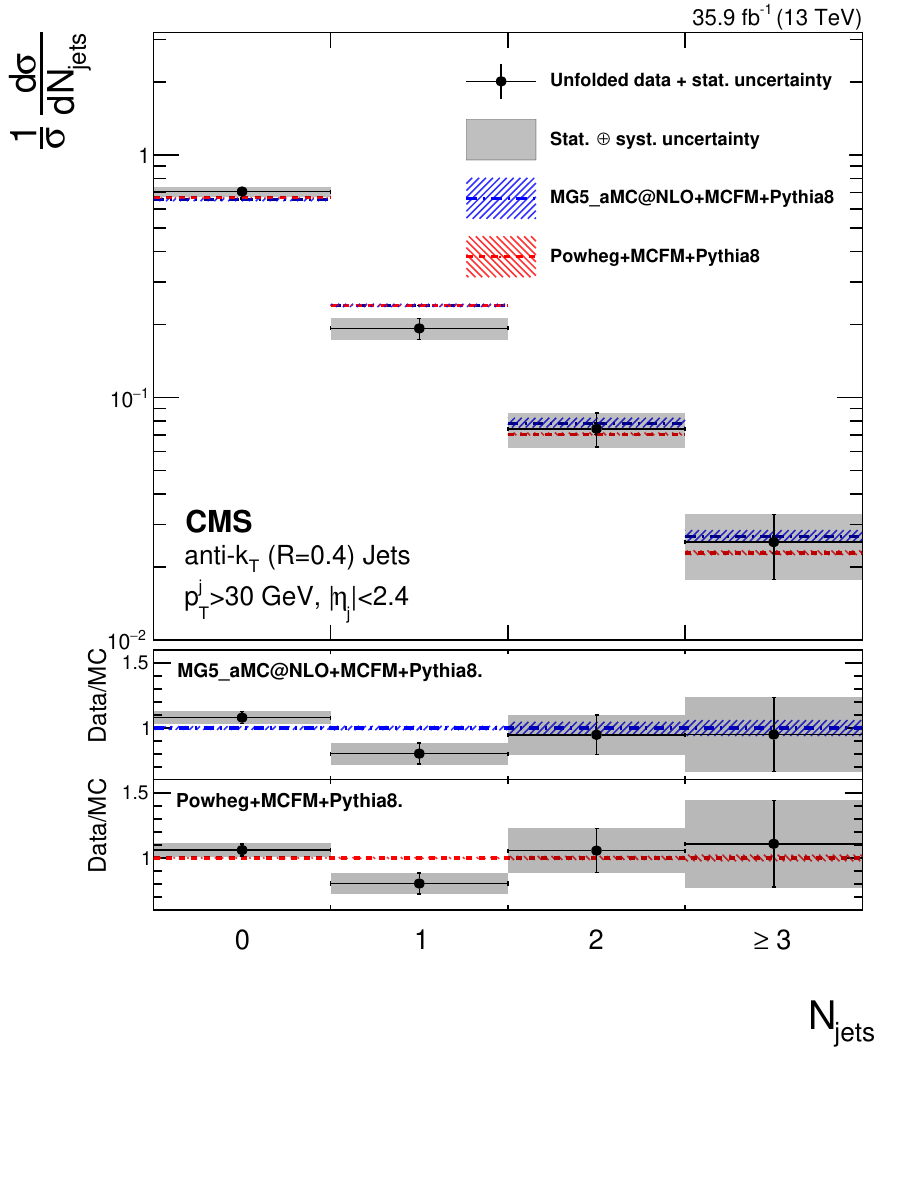}

   \caption{Differential cross sections normalized to the cross section of $\Pp\Pp\to\PZ\PZ\to 4\ell$ as a function of the multiplicity of jets with $\aetaj<4.7$ (top panels) and $\aetaj<2.4$ (bottom panels), for the 8 (left) and 13 (right)\TeV data.
   Other details are as described in the caption of Fig.~\ref{fig:diff_xs_nJets_notnorm}.
   }
   \label{fig:diff_xs_nJets}
\end{figure*}

\begin{figure*}[hbtp]
  \centering
    \includegraphics[width=0.95\cmsFigWidth]{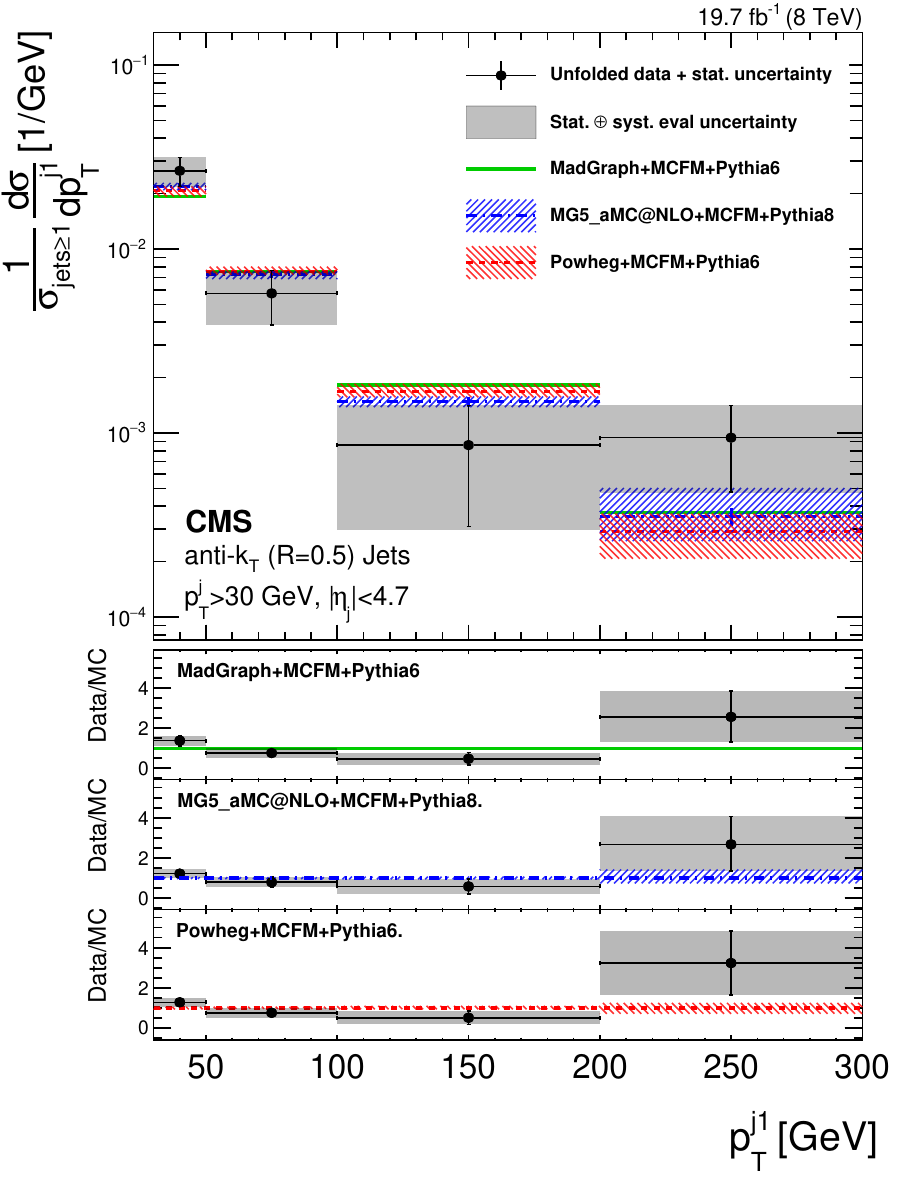}
    \includegraphics[width=0.95\cmsFigWidth]{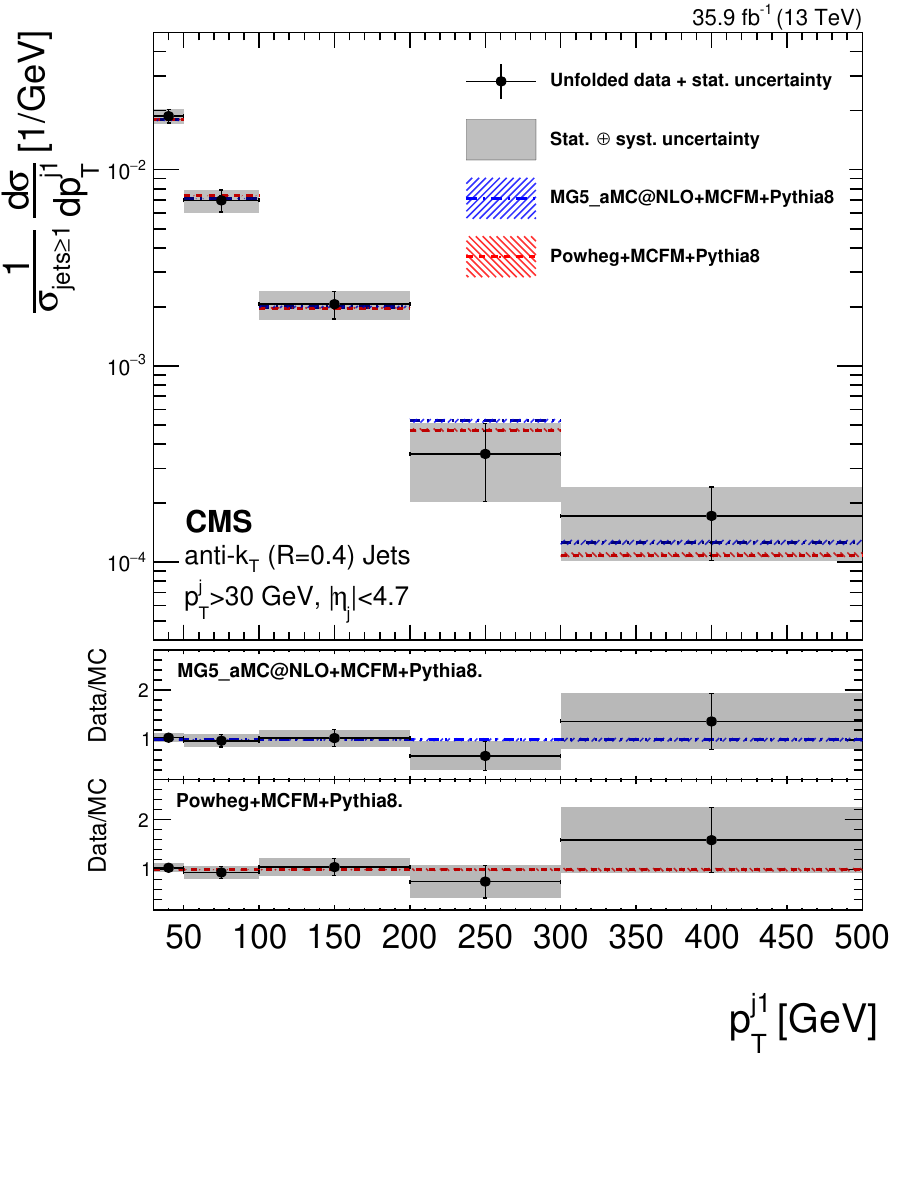}
    \includegraphics[width=0.95\cmsFigWidth]{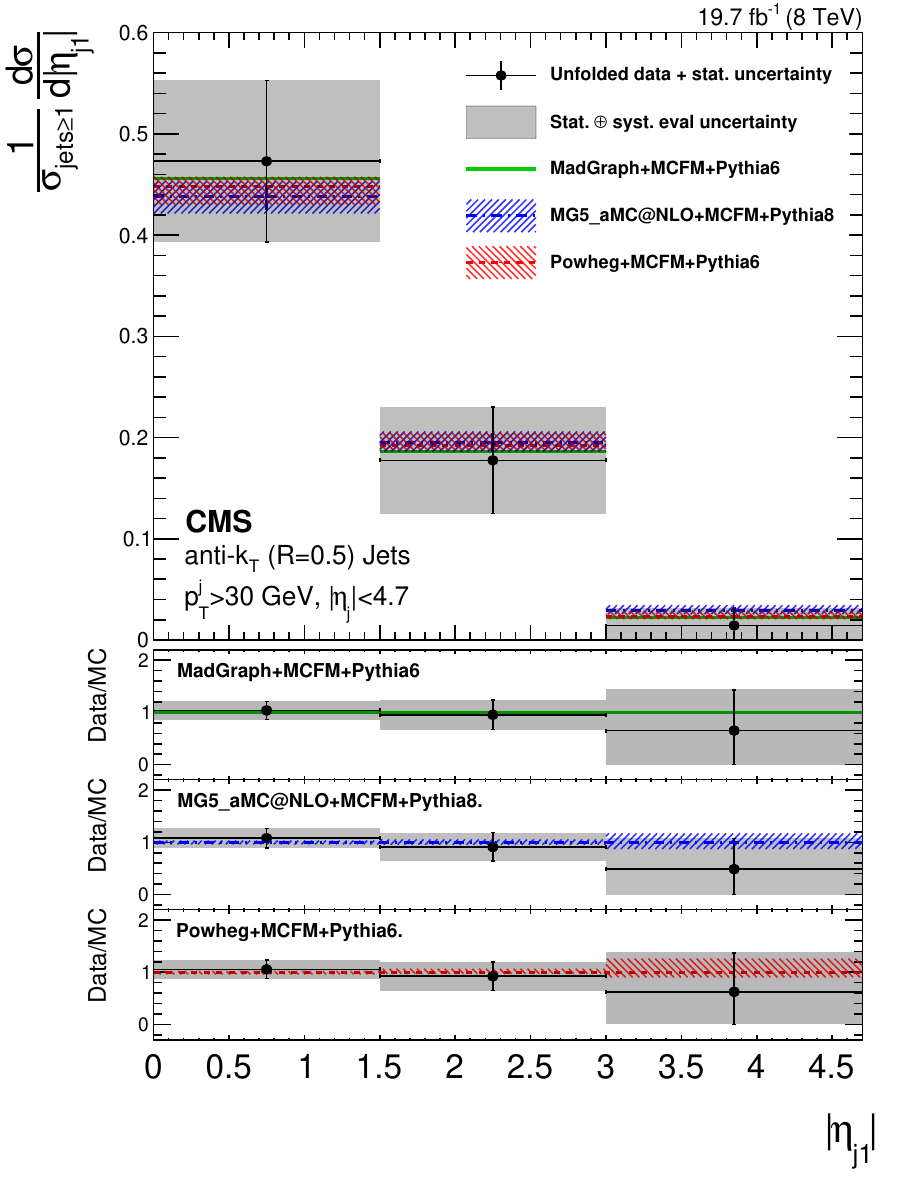}
    \includegraphics[width=0.95\cmsFigWidth]{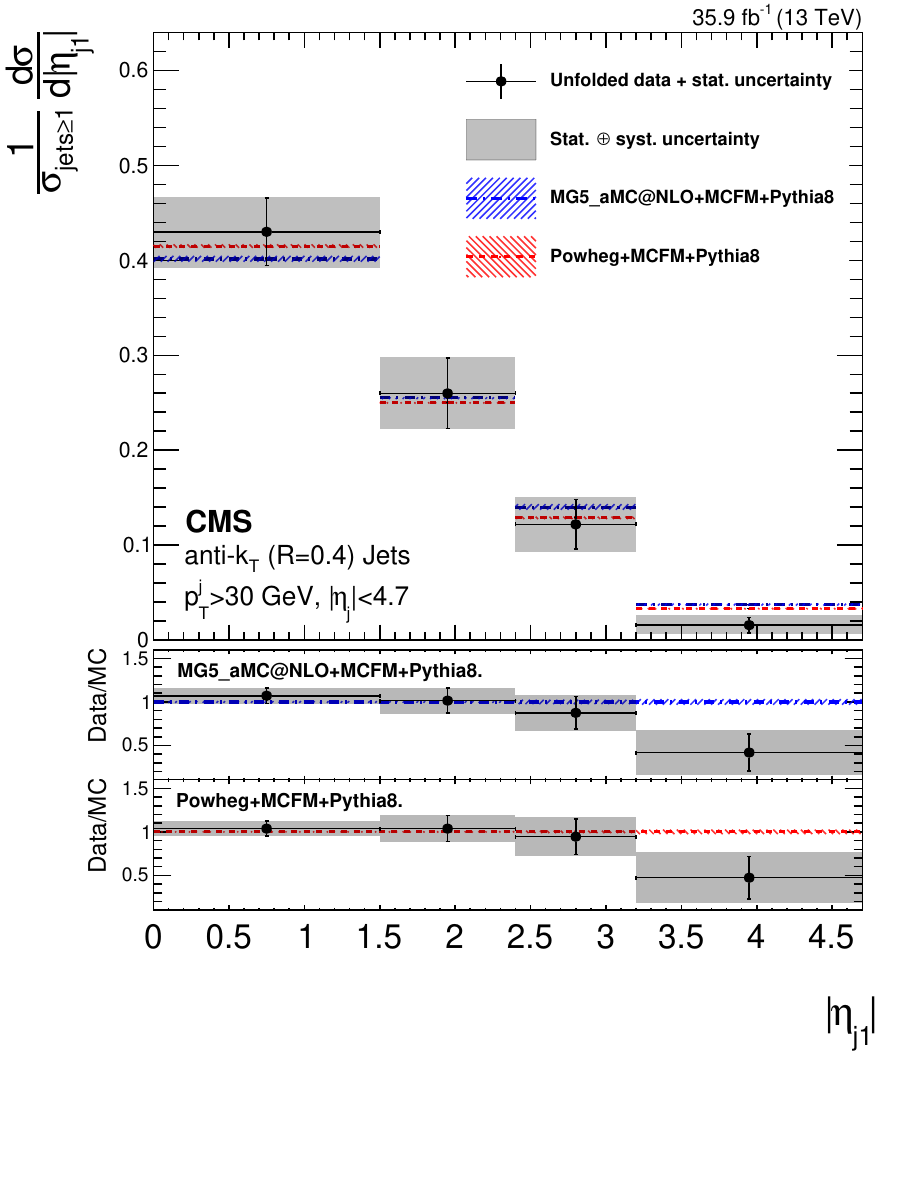}
   \caption{Differential cross sections normalized to the cross section for $\Njets \ge 1$ of $\Pp\Pp\to\PZ\PZ\to 4\ell$ as a function of the \pt-leading jet transverse momentum (top panels) and the absolute value of the pseudorapidity (bottom panels), for the 8 (left) and 13 (right)\TeV data.
   Other details are as described in the caption of Fig.~\ref{fig:diff_xs_nJets_notnorm}.
   }
   \label{fig:diff_xs_jet1}
\end{figure*}

\begin{figure*}[!hbtp]
  \centering
    \includegraphics[width=0.95\cmsFigWidth]{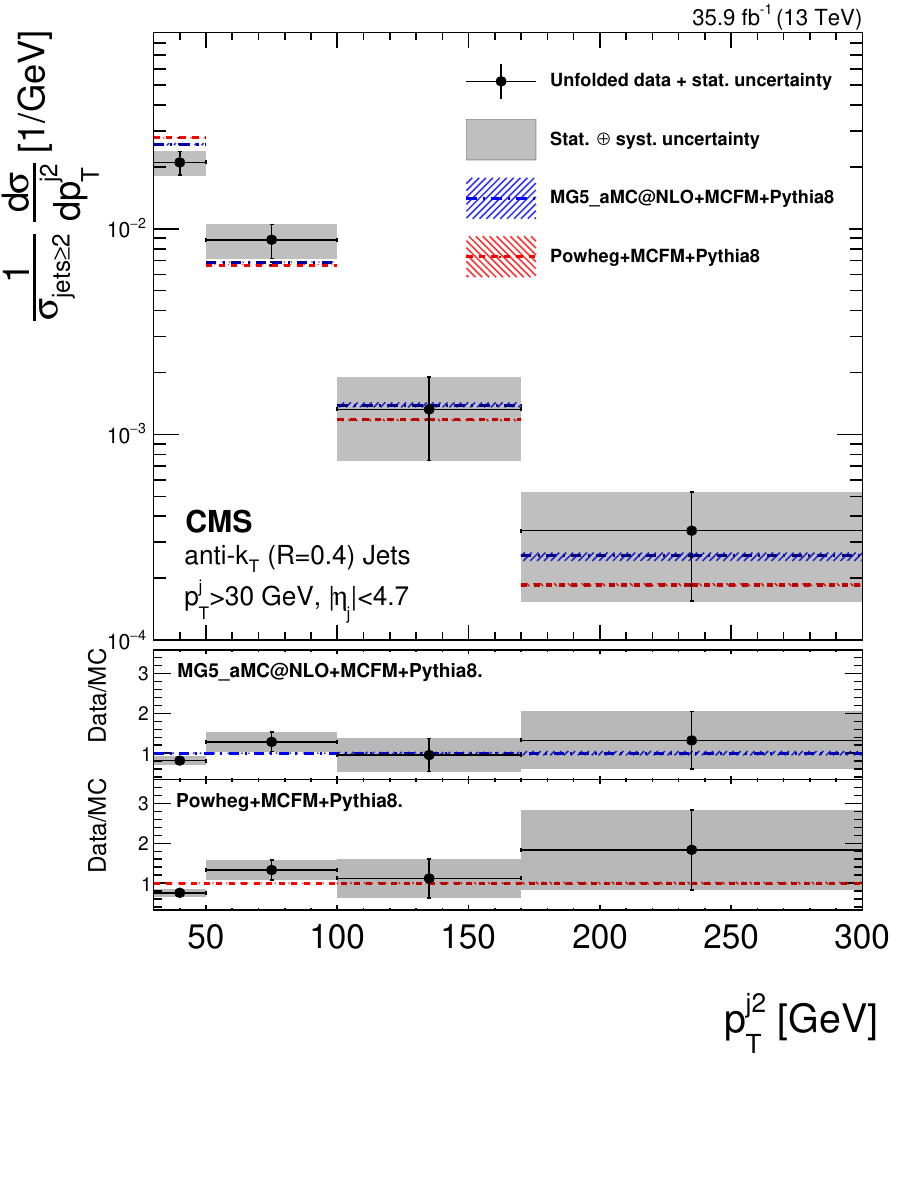}
    \includegraphics[width=0.95\cmsFigWidth]{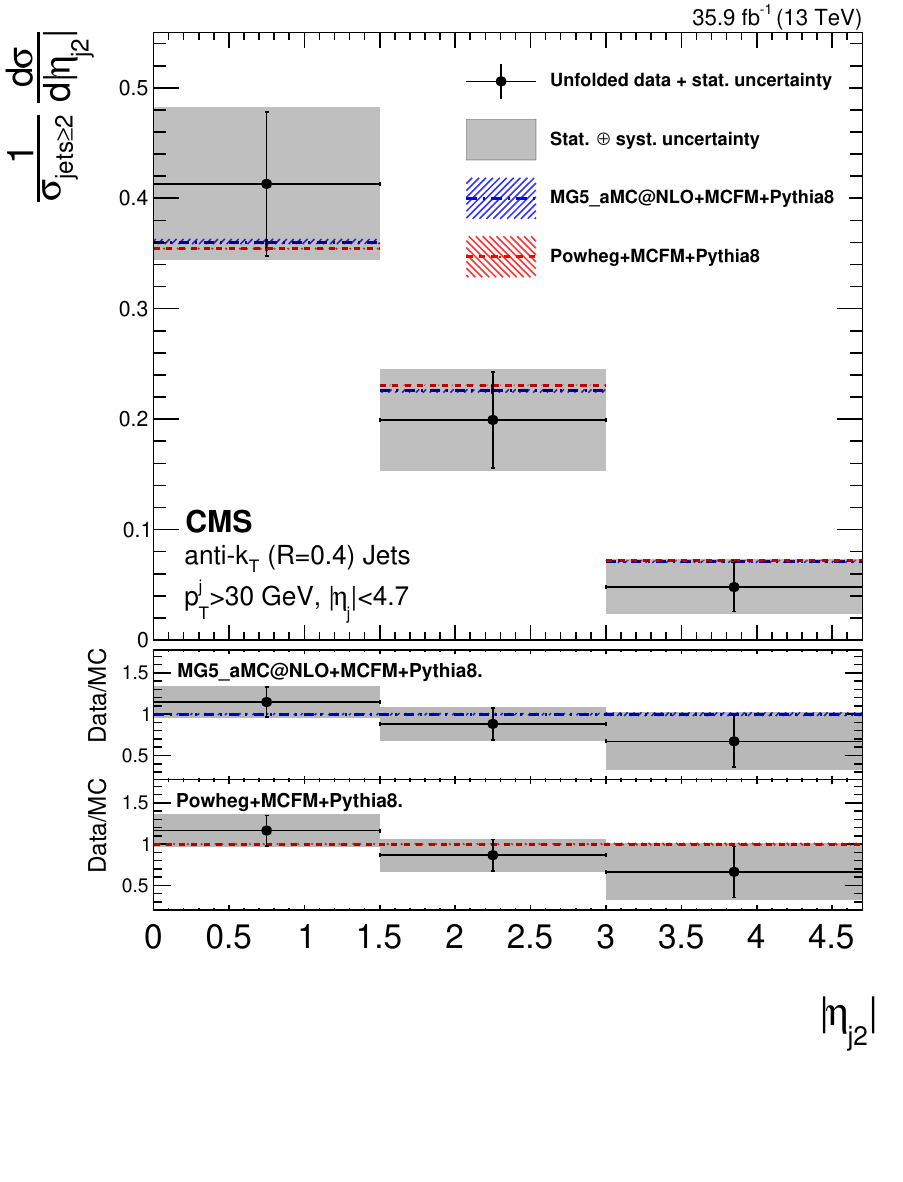}
   \caption{Differential cross sections normalized to the cross section for $\Njets \ge 2$ of $\Pp\Pp\to\PZ\PZ\to 4\ell$ at $\sqrt{s} = 13\TeV$  as a function of the \pt-subleading jet transverse momentum (left) and the absolute value of the pseudorapidity (right).
    Other details are as described in the caption of Fig.~\ref{fig:diff_xs_nJets_notnorm}.
   }
   \label{fig:diff_xs_jet2}
\end{figure*}

\begin{figure*}[hbtp]
  \centering
    \includegraphics[width=0.95\cmsFigWidth]{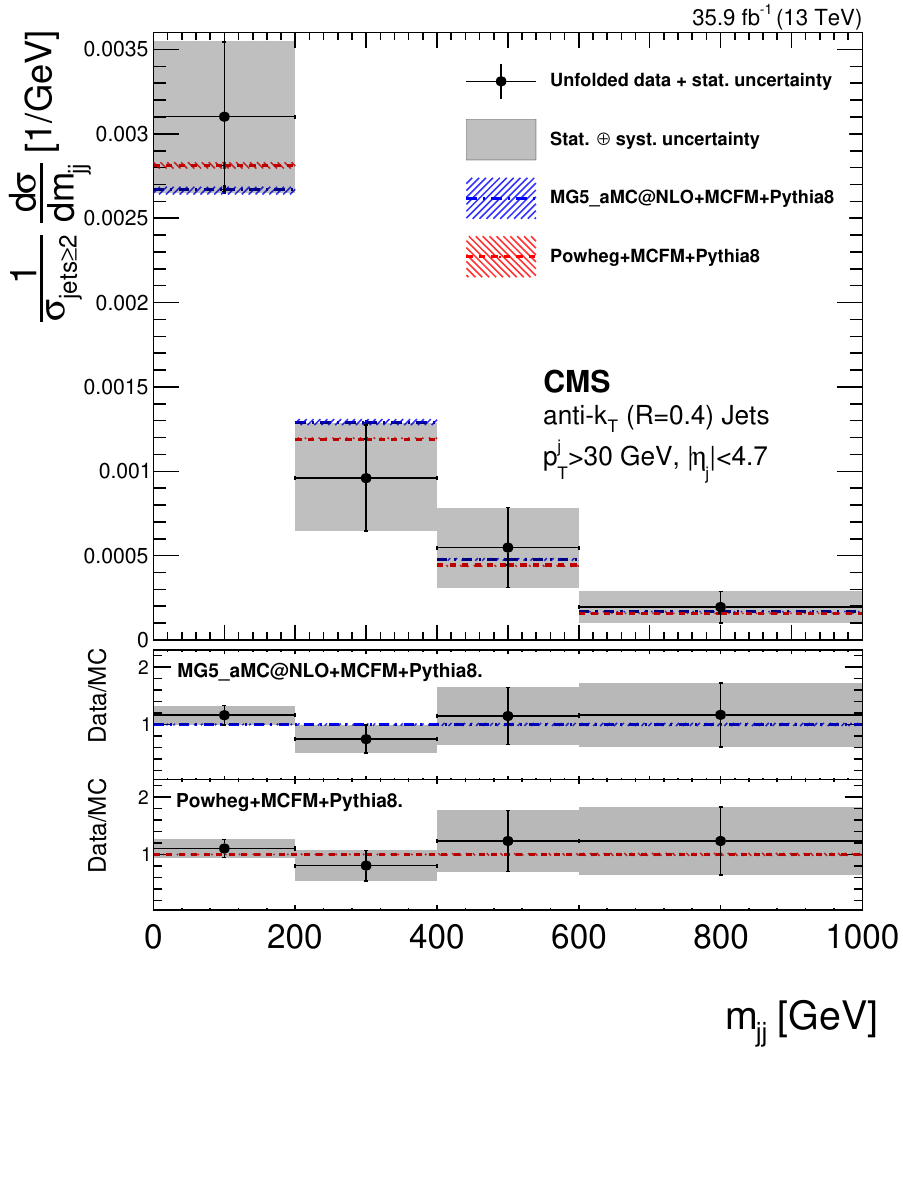}
    \includegraphics[width=0.95\cmsFigWidth]{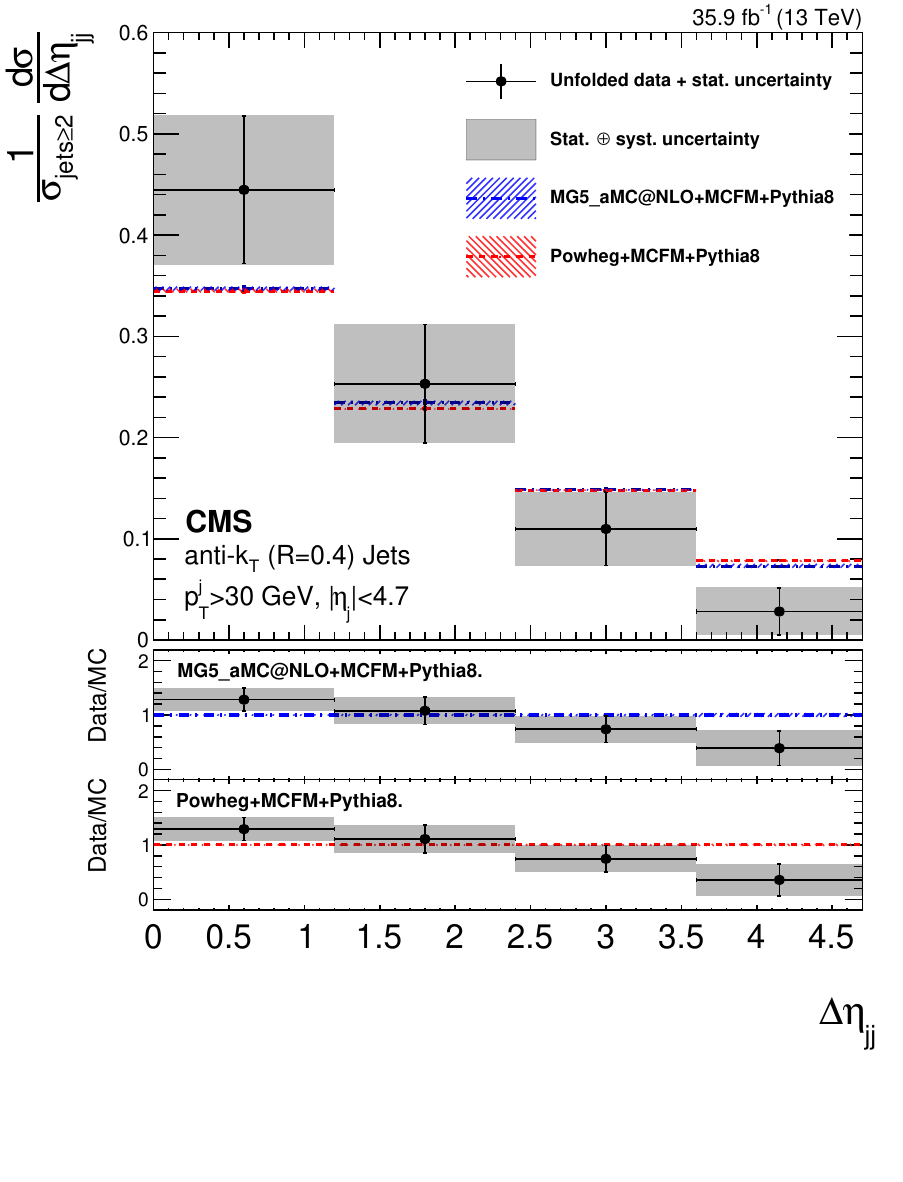}

   \caption{Differential cross sections normalized to the cross section for $\Njets \ge 2$ of $\Pp\Pp\to\PZ\PZ\to 4\ell$ at $\sqrt{s} = 13\TeV$  as a function of the invariant mass of the two \pt-leading jets (left) and their pseudorapidity separation (right).
    Other details are as described in the caption of Fig.~\ref{fig:diff_xs_nJets_notnorm}.
   }

  \label{fig:diff_xs_mjj_deta}
\end{figure*}

The differential distributions, normalized to the cross sections, are presented in Figs.~\ref{fig:diff_xs_nJets}--\ref{fig:diff_xs_mjj_deta} together
with the theoretical predictions. For the theoretical predictions, only the uncertainty in the shape is included, which yields a smaller uncertainty compared to the unnormalized case.
Figure~\ref{fig:diff_xs_nJets} (top panels) shows the normalized differential cross section as a function of the jet multiplicity, with $\aetaj < 4.7$.
The observed fraction of events in the first bin with zero jets is larger than the predicted value, while for 1, 2, and $\geq$ 3 jets, the fraction is lower.
Better agreement is observed  for $\aetaj<2.4$ (Fig.~\ref{fig:diff_xs_nJets}, bottom panels).
The measurements of the differential cross section as a function of the jet multiplicity are fairly well reproduced by the predictions both at 8 and 13\TeV
when NLO matrix-element calculations are used in conjunction with \PYTHIA 8 for parton showering, hadronization, and underlying event simulation.
In the data, jets tend to have a lower \pt value than in the simulations and therefore, on average, they are less
likely to pass the 30\GeV threshold, thus increasing the number of events with no jets. The observation of fewer events
than expected with at least one jet can be ascribed to a softer distribution of the transverse momentum of the
hadronic particles recoiling against the diboson system. This explanation is supported by the measurement of a softer-than-expected
\pt distribution of the $\PZ\PZ$ system~\cite{ZZXS8TeVPaper,ZZXS13TeVPaper}. The observed discrepancy may be due to higher-order
corrections to $\PZ\PZ$ production, not included in MC samples used in this analysis, or to the parton shower modeling.

Figure~\ref{fig:diff_xs_jet1} shows the differential cross sections at 8 and 13\TeV as functions of the transverse momentum and pseudorapidity of the \pt-leading jet, normalized to the cross section for $\Njets \ge 1$.
Figures~\ref{fig:diff_xs_jet2} and \ref{fig:diff_xs_mjj_deta} show the cross section at 13\TeV as a function of several variables for events with $\Njets \ge 2$, normalized to the corresponding cross section.
More specifically, Fig. \ref{fig:diff_xs_jet2} presents the normalized differential cross sections as functions of the transverse momentum and pseudorapidity of the \pt-subleading jet,
while Fig.~\ref{fig:diff_xs_mjj_deta} displays the differential cross section as a function of \mjj and \detajj.

Overall agreement is observed between data and theoretical predictions for all measurements related to the \pt-leading and subleading jets. The \detajj distribution (Fig.~\ref{fig:diff_xs_mjj_deta}, right) measured with 13\TeV data tends to be steeper than the MC predictions, but the differences are not statistically significant.

\section{Summary}
\label{sec:summary}

The differential cross sections for the production of \PZ pairs in the four-lepton final state in association with
jets in proton-proton collisions at $\sqrt{s}=8$ and 13\TeV have been measured.
The data correspond to an integrated luminosity of 19.7 (35.9)\fbinv for a center-of-mass energy of 8~(13)\TeV.
Cross sections are presented for the production of a pair of \PZ bosons
as a function of the number of jets, the transverse momentum \pt, and
pseudorapidity of the \pt-leading and subleading jets. Distributions of the
invariant mass of the two \pt-leading jets and their separation in pseudorapidity
are also presented.
Good agreement is observed between the measurements and the theoretical predictions when next-to-leading order matrix-element calculations are
used together with the \PYTHIA parton shower simulation.
Cross sections for $\PZ\PZ$ production in association with jet have been measured with a precision ranging from 10 to 72\%  (8 to 38\%) at 8 (13)\TeV, for jet
multiplicities ranging from 0 to $\geq$ 3.
The systematic uncertainty is of the same size, or smaller, than the statistical one.
Analyses using future, larger data sets, with smaller statistical uncertainties, will allow the theoretical prediction of \zzjets to undergo more stringent tests.

\begin{acknowledgments}

We congratulate our colleagues in the CERN accelerator departments for the excellent performance of the LHC and thank the technical and administrative staffs at CERN and at other CMS institutes for their contributions to the success of the CMS effort. In addition, we gratefully acknowledge the computing centers and personnel of the Worldwide LHC Computing Grid for delivering so effectively the computing infrastructure essential to our analyses. Finally, we acknowledge the enduring support for the construction and operation of the LHC and the CMS detector provided by the following funding agencies: BMWFW and FWF (Austria); FNRS and FWO (Belgium); CNPq, CAPES, FAPERJ, and FAPESP (Brazil); MES (Bulgaria); CERN; CAS, MoST, and NSFC (China); COLCIENCIAS (Colombia); MSES and CSF (Croatia); RPF (Cyprus); SENESCYT (Ecuador); MoER, ERC IUT, and ERDF (Estonia); Academy of Finland, MEC, and HIP (Finland); CEA and CNRS/IN2P3 (France); BMBF, DFG, and HGF (Germany); GSRT (Greece); NKFIA (Hungary); DAE and DST (India); IPM (Iran); SFI (Ireland); INFN (Italy); MSIP and NRF (Republic of Korea); LAS (Lithuania); MOE and UM (Malaysia); BUAP, CINVESTAV, CONACYT, LNS, SEP, and UASLP-FAI (Mexico); MBIE (New Zealand); PAEC (Pakistan); MSHE and NSC (Poland); FCT (Portugal); JINR (Dubna); MON, RosAtom, RAS and RFBR (Russia); MESTD (Serbia); SEIDI, CPAN, PCTI and FEDER (Spain); Swiss Funding Agencies (Switzerland); MST (Taipei); ThEPCenter, IPST, STAR, and NSTDA (Thailand); TUBITAK and TAEK (Turkey); NASU and SFFR (Ukraine); STFC (United Kingdom); DOE and NSF (USA).

\hyphenation{Rachada-pisek} Individuals have received support from the Marie-Curie program and the European Research Council and Horizon 2020 Grant, contract No. 675440 (European Union); the Leventis Foundation; the A. P. Sloan Foundation; the Alexander von Humboldt Foundation; the Belgian Federal Science Policy Office; the Fonds pour la Formation \`a la Recherche dans l'Industrie et dans l'Agriculture (FRIA-Belgium); the Agentschap voor Innovatie door Wetenschap en Technologie (IWT-Belgium); the F.R.S.-FNRS and FWO (Belgium) under the ``Excellence of Science - EOS" - be.h project n. 30820817; the Ministry of Education, Youth and Sports (MEYS) of the Czech Republic; the Lend\"ulet (``Momentum") Program and the J\'anos Bolyai Research Scholarship of the Hungarian Academy of Sciences, the New National Excellence Program \'UNKP, the NKFIA research grants 123842, 123959, 124845, 124850 and 125105 (Hungary); the Council of Science and Industrial Research, India; the HOMING PLUS program of the Foundation for Polish Science, cofinanced from European Union, Regional Development Fund, the Mobility Plus program of the Ministry of Science and Higher Education, the National Science Center (Poland), contracts Harmonia 2014/14/M/ST2/00428, Opus 2014/13/B/ST2/02543, 2014/15/B/ST2/03998, and 2015/19/B/ST2/02861, Sonata-bis 2012/07/E/ST2/01406; the National Priorities Research Program by Qatar National Research Fund; the Programa Estatal de Fomento de la Investigaci{\'o}n Cient{\'i}fica y T{\'e}cnica de Excelencia Mar\'{\i}a de Maeztu, grant MDM-2015-0509 and the Programa Severo Ochoa del Principado de Asturias; the Thalis and Aristeia programs cofinanced by EU-ESF and the Greek NSRF; the Rachadapisek Sompot Fund for Postdoctoral Fellowship, Chulalongkorn University and the Chulalongkorn Academic into Its 2nd Century Project Advancement Project (Thailand); the Welch Foundation, contract C-1845; and the Weston Havens Foundation (USA).

\end{acknowledgments}

\bibliography{auto_generated}
\cleardoublepage \appendix\section{The CMS Collaboration \label{app:collab}}\begin{sloppypar}\hyphenpenalty=5000\widowpenalty=500\clubpenalty=5000\input{SMP-17-005-authorlist.tex}\end{sloppypar}
\end{document}

%% file: SMP-17-005-authorlist.tex
\vskip\cmsinstskip
\textbf{Yerevan Physics Institute, Yerevan, Armenia}\\*[0pt]
A.M.~Sirunyan, A.~Tumasyan
\vskip\cmsinstskip
\textbf{Institut f\"{u}r Hochenergiephysik, Wien, Austria}\\*[0pt]
W.~Adam, F.~Ambrogi, E.~Asilar, T.~Bergauer, J.~Brandstetter, M.~Dragicevic, J.~Er\"{o}, A.~Escalante~Del~Valle, M.~Flechl, R.~Fr\"{u}hwirth\cmsAuthorMark{1}, V.M.~Ghete, J.~Hrubec, M.~Jeitler\cmsAuthorMark{1}, N.~Krammer, I.~Kr\"{a}tschmer, D.~Liko, T.~Madlener, I.~Mikulec, N.~Rad, H.~Rohringer, J.~Schieck\cmsAuthorMark{1}, R.~Sch\"{o}fbeck, M.~Spanring, D.~Spitzbart, A.~Taurok, W.~Waltenberger, J.~Wittmann, C.-E.~Wulz\cmsAuthorMark{1}, M.~Zarucki
% ...
\vskip\cmsinstskip
% ...
1:  Also at Vienna University of Technology, Vienna, Austria\\
% ...
